\documentclass{article}
{\usepackage[utf8]{inputenc}}           
\usepackage[USenglish]{babel}			 
\usepackage[final]{pdfpages}
\usepackage{multirow}
\usepackage{ragged2e}
\usepackage{amsmath}
\usepackage{commath}
\usepackage{tikz}
\usepackage{pgfplots}
\usepackage{color}
\usepackage{comment}
\usepackage{todonotes}
\usepackage{subfigure}
\usepackage[export]{adjustbox}
\pgfplotsset{compat=newest}
\usepackage{gensymb}
\usetikzlibrary{arrows,decorations,backgrounds,patterns,matrix,shapes,fit,calc,shadows,plotmarks,pgfplots.groupplots}

\begin{document}
\vspace{-10cm}
\title{Transition radiation based transverse beam diagnostics for non relativistic ion beams}
\author{R. Singh\thanks{r.singh@gsi.de}\textsuperscript{1}, T. Reichert\textsuperscript{1} and B. Walasek-Hoehne\textsuperscript{1} \\
	    \textsuperscript{1}GSI Helmholtzzentrum f\"ur Schwerionenforschung GmbH,\\ Darmstadt, Germany  \\}
	 
\maketitle

\begin{abstract}
The usage of optical transition radiation for profile monitoring of relativistic electron beams is well known. This report presents the case for beam diagnostic application of optical transition radiation for non-relativistic ion beams. The angular distribution of the transition radiation emitted from few target materials for ion beam irradiation is shown. In addition to expected linearly polarized transition radiation in the plane of observation, a large amount ($\approx \times 20$) of unpolarized radiation is observed increasing towards grazing angles. The unpolarized radiation has the characteristics of transition radiation and is understood as the transition radiation generated from a "strongly" rough target surface. This increase in amount of radiation towards the detector can be used advantageously towards transverse profile measurements and potentially other beam parameters. Further systematic effects such as the dependence of light yield on beam current, comparison of the measured transverse profiles with Secondary electron emission based grid (SEM grid), target heating etc. are also shown.
\end{abstract}

\section{Introduction}
Transition radiation is generated over a wide frequency range when a charged particle traverses two different media. The existence of transition radiation was first predicted by Ginzburg and Frank~\cite{Ginzburg} where the expressions for the radiated spectral intensity with the far field angular distribution were derived. The radiation intensity is typically evaluated between two homogeneous media with different permittivities such as the vacuum-metal boundary which is often the case for most practical realizations.

 A common picture for visualizing transition radiation for the case of a planar vacuum-metal interface is the following. As the charged particle approaches the interface from the vacuum side, the coulomb fields associated with the charge are terminated on the metal such that the boundary conditions are satisfied by the induced polarization.
 However, when the charge hits the metal vacuum-metal interface, the boundary conditions can only be satisfied with addition of 
 radiating electromagnetic fields which is referred to as transition radiation~\cite{Ginzburg1}.
 A similar process happens on appearance of the charged particle through the metal sheet on the other side at the metal-vacuum interface. The emitted radiation is often separated as backward transition radiation (BTR) when the emission is in the first medium and forward transition radiation (FTR) when it is in the second medium. For non-relativistic ion beams, the beam is typically deposited into the first few micrometers of the target and there is no forward transition radiation expected unless very thin foils are used. One has to note, that although simple descriptions for transition radiation are given for vacuum-metal interface, transition radiation occurs at every material interface or by traversing an inhomogeneous medium and is purely a surface phenomenon.
 With respect to particle bunches another distinction commonly used in the jargon of transition radiation is the coherent against incoherent transition radiation. The radiation is referred to as coherent if the observed frequency range of the radiation is contained in the frequency range of the source charge distribution. Coherent transition radiation for optical frequencies would only occur for bunches with temporal lengths in the order of few femto seconds (fs).

\subsection{Optical transition radiation from smooth target.}

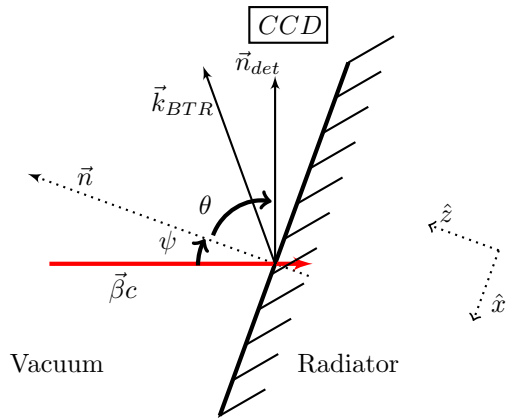
\begin{figure}[ht]
\centering
\begin{tikzpicture}

\draw[ultra thick] (1,5)--++(250:5.0);
\draw[thick,dotted,latex'-] (1,5)++(250:5.7/2)++(160:3.5)--++(340:4.0);
\draw[ultra thick,latex'-,red] (1,5)++(250:5.7/2)++(0:0.5)--++(180:3.5);
\draw[thick,-latex'] (1,5)++(250:5.7/2)--++(90:2.5);
\draw[thick,-latex'] (1,5)++(250:5.7/2)--++(110:2.8);
\newcommand\shadeang{30}
\newcommand\shadelen{0.7}
\foreach[count=\i, evaluate=\i as \x using (\i*0.5)] \y in {0,1,...,10}
{
\draw[thick] (1,5)--++(250:\x-0.5)--++(\shadeang:\shadelen);
}

\draw[thick,dotted, ->] (3,2.5)--++(250:1.0);
\draw[thick,dotted, ->] (3,2.5)--++(160:1.0);

\draw[ultra thick, ->] (-1.0,2.3) arc (180:150:0.7);
\draw[ultra thick, ->] (-0.8,2.7) arc (160:80:0.7);
\node[thick,text width=1.0cm,align=center] (Medium1) at (-3.0,1.0) {Vacuum};
\node[thick,text width=2.0cm,align=center] (Plate) at (1.0,1.0) {Radiator};
\node[thick,text width=4.0cm,align=center] (n) at (-2.5,3.5) {$\vec{n}$};
\node[thick,text width=4.0cm,align=center] (kbtr) at (-1.2,4.5) {$\vec{k}_{BTR}$};
\node[thick,text width=0.8cm,draw,align=center] (cam) at (0.2,5.5) {$CCD$};
\node[thick,text width=1.0cm,align=center] (camvec) at (-0.2,5.0) {$\vec{n}_{det}$};
\node[thick,text width=4.0cm,align=center] (bc) at (-2.0,2) {$\vec{\beta}c$};
\node[thick,text width=4.0cm,align=center] (psi) at (-1.4,2.6) {$\psi$};
\node[thick,text width=4.0cm,align=center] (theta) at (-0.9,3.1) {$\theta$};
\node[thick,text width=4.0cm,align=center] (x) at (3.0,1.8) {$\hat{x}$};
\node[thick,text width=4.0cm,align=center] (x) at (2.3,3.0) {$\hat{z}$};

\end{tikzpicture}
\caption{Charged particle beam impinging on a metal target \label{fig1}}
\end{figure}
Figure~\ref{fig1} shows the schematic of the particle trajectory, target or radiator and the detector along with the relevant symbols. The co-ordinate system is chosen in line with~\cite{TM}, i.e. the OTR target plane is defined as the $x-y$ plane and the target normal is towards the $z$ axis. The plane of incidence consists of target normal $\vec{n}$ and beam incidence vector $\frac{\vec{\beta}}{|\beta|}$ and the angle between them ($\psi$) is referred to as the irradiation angle. The plane of incidence, in the shown case co-incides with the $x-z$ plane. The plane of radiation consists of $\vec{n}$ and radiation wavevector $\vec{k_{BTR}}$ while the plane of observation contains beam incidence vector $\vec{\beta}$ and the detector position $\vec{n_{det}}$.

 In our designed experimental set-up with a smooth target, the beam incidence vector, the target normal and detector position vector are in the same plane, i.e. the plane of incidence and plane of observation co-incide. Further, since only the radiation in the plane of observation is detected due to a small acceptance of the detector (i.e. $\frac{\vec{k_{BTR}}}{|k_{BTR}|} \approx \vec{n_{det}}$), the relevant radiation vector also lies in the same plane as incidence and observation plane.

If an ion beam with a charge state $Z$ is traversing from medium 1 with $\epsilon_1 = 1$ into a medium 2 with arbitrary relative permittivity $\epsilon_2=\epsilon$ at an irradiation angle $\psi$, the angular distribution of the far field spectral radiant intensity with a polarization parallel to plane of radiation ($p$ polarized) as a function of irradiation angle and emission angle $\theta$ is given as~\cite{TM},

\begin{equation}\label{eq:tr}
\frac{dI_\parallel(n,\omega,\theta,\phi,\psi)}{d\Omega d\omega}= \frac{Z^2e^2\beta_z^{2}\cot^2{\theta}|\epsilon-1|^2}{4\pi^3\epsilon_0 c[(1-\beta_x\cos{\theta_x})^2-\beta_z^2\cos{\theta}]^2}\|A||^2 
\end{equation}
where, 
\begin{equation}
\begin{aligned}
A&=\frac{(1+\beta_z\sqrt{\epsilon-\sin^2{\theta}}-\beta_z^2-\beta_x\cos{\theta_x})\sin^2{\theta}-\beta_x\beta_z\cos{\theta_x}\sqrt{\epsilon-\sin^2{\theta}}}{(\epsilon\cos{\theta} + \sqrt{\epsilon-\sin^2{\theta}})(1-\beta_x\cos{\theta_x}+\beta_z\sqrt{\epsilon-\sin^2{\theta}})}
\end{aligned}
\end{equation}
and, $\beta_x = \beta\sin{\psi}$, $\beta_z= \beta\cos{\psi}$ and $\cos{\theta_x}= \sin{\theta}\cos{\phi}$.

\begin{figure}[h]
\centering  
\subfigure[]{\includegraphics[width=0.4\textwidth]{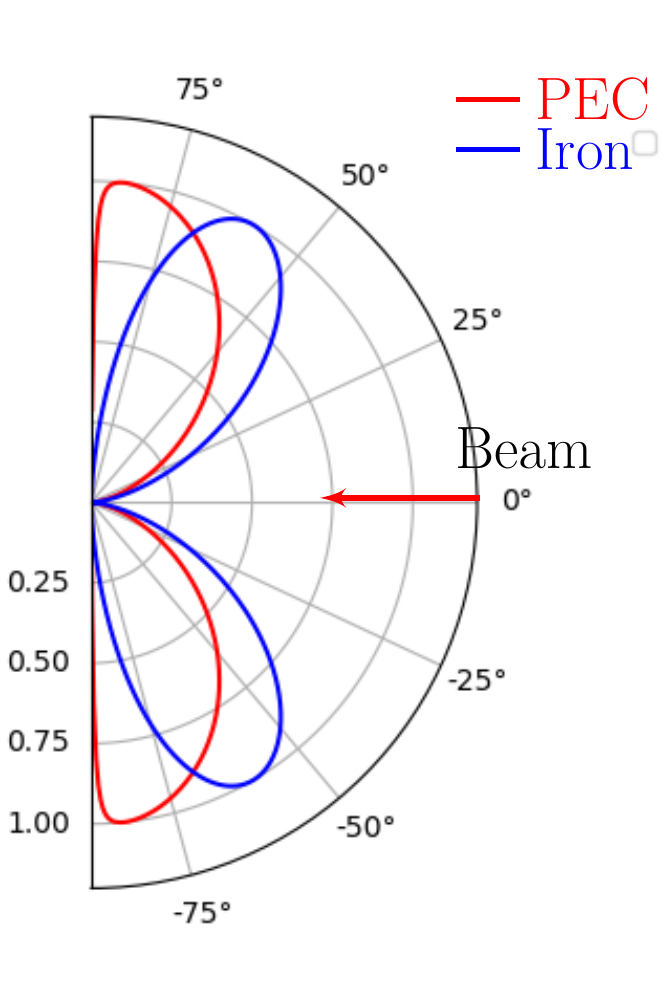}}
\subfigure[]{\includegraphics[width=0.58\textwidth]{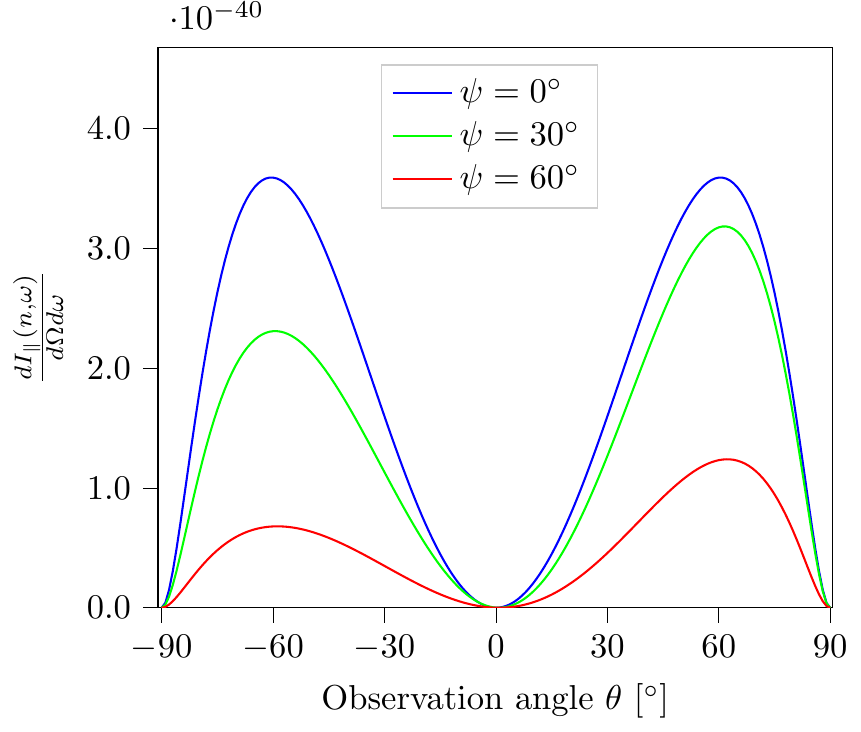}} 
    
\caption{(a) Polar plot showing the angular distribution of transition radiation emission from an Iron target in comparison to perfect electric conductor (PEC) when irradiated by a singly charged particle with velocity $\beta = 0.15$ at $\psi = 0$. (b) The angular distribution of spectral intensity for an Iron target when irradiated at different angles.}
    \label{fig:ang_dist}
\end{figure}
There is also a component polarized perpendicular to the plane of radiation (s-polarized) $I_\perp$, however $I_\perp \approx I_\parallel\cdot\beta^4$ 
It is thus expected to be negligible in comparison to parallel component for lower energies relevant for our studies $\beta < 0.5$.
For smooth targets oriented as shown in Fig.~\ref{fig1}, we expect the radiation detected  by the camera to be linearly polarized since the target normal ($\vec{n}$) lies in the plane of observation.
For rough targets, the situation can change drastically since the target normal varies across the whole target depending on the specifics of the target surface structure and the polarization of the emitted radiation will change correspondingly. This also has consequences for the angular distribution as discussed in next subsection.
 Fig~\ref{fig:ang_dist}(a) shows a polar diagram representing the angular distribution given by Eq.~\ref{eq:tr} for a unit charge with velocity $\beta = 0.15$ incident normally on perfect electrical conductor (PEC) and a realistic conductor i.e. Iron (Fe). Fig~\ref{fig:ang_dist}(b) shows the angular distribution for various irradiation angles. 
 Here the relative permittivity of Iron at $500$ nm was used which can be found in~\cite{Christy}.
 The peak of the radiation for an iron target is at $\theta ~\approx 60 \degree$ irrespective of the angle of irradiation $\psi$ while for PEC, it is $\psi = 90\degree$. The cumulative spectral intensity is highest for normal incidence onto the target.

With this smooth surface assumption, optical transition radiation from low electron beams have been already been observed~\cite{Bal,Fiorito}. For low energy hadron beams, usage of OTR for diagnostic purposes was proposed in 2008~\cite{Lumpkin}. Following which some pilot measurements were done, which confirmed the charge state dependency of light yield. Spectroscopic investigations showed a broadband radiation indicating the presence of OTR were reported~\cite{Walasek}. Polarization, angular dependence and light yield were not studied in detail. Since the charge state dependency also exists for other light producing mechanisms like beam induced fluorescence (BIF), the data did not definitively confirm the measured radiation as OTR in those pilot studies. This work tries to build upon that experience.

\subsection{Optical transition radiation from a rough target}\label{sec:roughtarget}

An important concept for transition radiation generation is the "effective source size" $r_{eff}$ which for a certain radiation wavelength $\lambda$ for a beam corresponding to $\beta$ is given as $\beta\gamma\lambda$~\cite{Verzilov}. It is clear that for lower betas $\beta<0.5$ and the radiation emitted in optical regime, it is smaller than a naturally unprocessed surface roughness. The targets used in this study were not controlled for surface roughness and thus expected to have an rms roughness of $\ge 500$ nm. This means that each structure on the  rough surface has the possibility to act as an individual radiator for transition and diffraction radiation with its own target normal which might not co-incide with plane of observation.
Another related concept is the so called formation zone~\cite{TM} and is given by, $L_{formation}=\frac{\lambda\beta\sqrt{\epsilon_0}}{2\pi (1-\beta\sqrt{\epsilon_0}\cos{\theta})}$ where $\epsilon_0$ is given as $(\epsilon_1+ \epsilon_2)/2$. Formation zone is comparable to the wavelength and should not play a role in optical regime for smooth targets with low energy beams. However, for rough targets, it can come into play where radiation can be different from shallow and deep grooves and could play a role for diffraction radiation.
\begin{figure}[h]
    \centering  
    \includegraphics[width=0.9\textwidth]{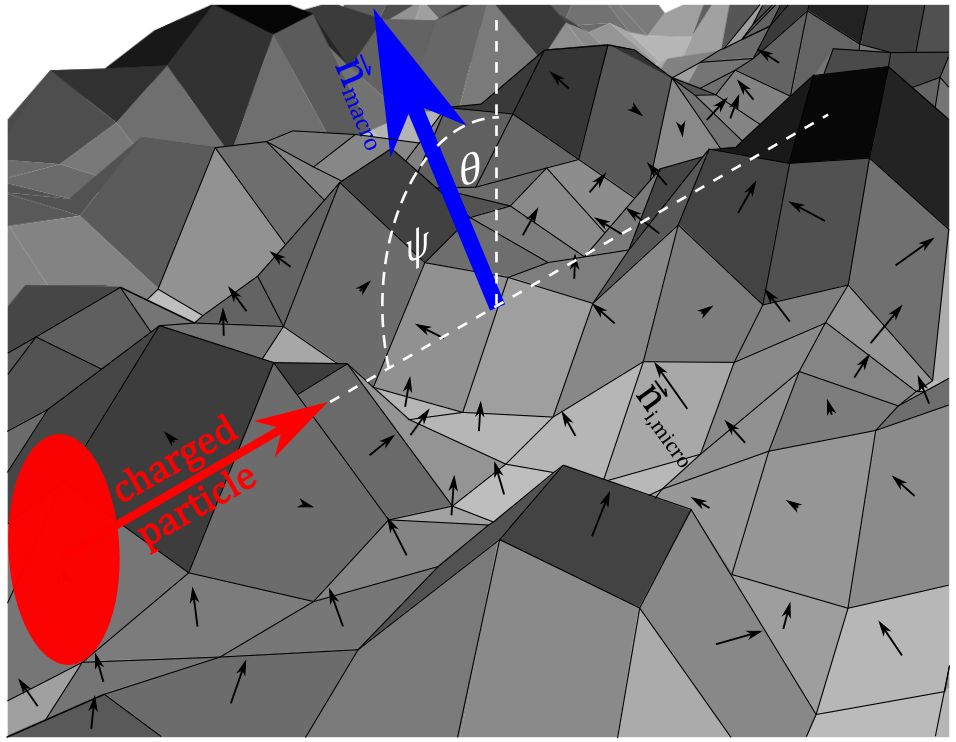}
        \caption{Schematic representation of a target with a rough surface hit by a charged particle under shallow incidence. Given the depicted piecewise planar micro-surfaces (with individual normals $\vec{n}_{i,micro}$) are bigger than the eff. transversal extent of the incident electric field ($r_\text{eff} = \beta \gamma\lambda$ in vacuum) they can be thought as individual TR sources}
    \label{fig2}
\end{figure}
The proportion of the polarized light measured in the plane of observation will thus be given by the relative distribution of microscopic target normals $\vec{n}_{micro}$ of individual radiators with respect to the macroscopic target normal $\vec{n}_{macro}$ which was earlier simply denoted as $\vec{n}$ (see Fig.~\ref{fig1}). Figure~\ref{fig2} shows a schematic of a rough target surface. We assume that the target normals for individual micro-radiators are randomly distributed in the 2D angular space with respect to $\vec{n_{macro}}$ and have no preferred direction, i.e. $\vec{n}_{micro}$ is a random variable with respect to $\theta,\phi$ with ${\bf E}[\vec{n}_{micro}] = \vec{n}_{macro}$ where {\bf E[]} is the expectation operator. However, only the subset of micro-radiators normals $(\vec{n}_{micro,OTR})$ matter for OTR generation, i.e. the ones which can be irradiated by the beam and ${\bf E}[\vec{n}_{micro,OTR}] = \vec{n}_{macro}+\alpha$. The value of $\alpha$ will depend on the exact nature of the surface structure, if we assume a simple step or groove type of structures on the surface, $\alpha \approx \pi/2 $. The red blob shows the field distribution of the charged particle and depicts the 
effective source size.

 A further complexity for transition radiation produced by low beta beams is the scattering of OTR light off the target surface and its diffraction at the edges of micro-radiators since a large part of the radiation is directed close to the target surface (see Fig.~\ref{fig:ang_dist}). 
 Scattering of light from a rough surface has been a subject of intensive study for many decades since all surfaces are rough to some extent. The studies have mainly been performed either from the material surface characterization perspective~\cite{Bickel} or were efforts towards building perfect mirrors~\cite{Ingers,Maradudin}. Broadly speaking, rough surfaces are typically characterized as "weak" or "strong" rough surfaces based on the "rms" roughness and characteristic length (correlation length) with respect to the incident light wavelength~\cite{Maradudin}. Strongly rough surfaces have characteristic lengths and rms roughness larger than the wavelength of interest. Further details on surface roughness and scattering can be found in these references and references therein~\cite{Ingers,Maradudin,Marvin_polarization}.
Together, the surface scattering and randomly distributed radiator angles especially for strong surface roughness can be of major consequence for the degree of polarization in the plane of observation as well as the angular distribution.

There has been early theoretical~\cite{Bagiyan} and experimental~\cite{Arutyunyan} work showing differences in optical transition radiation photon yield and polarization for low energy electron beams. These studies were mainly concerned with very shallow angles $\psi > 80 \degree$ and assigned the increase in radiation at shallow angles to diffraction radiation.
Further, it is known from literature that scattering via surface electromagnetic waves or surface polaritons is not expected to play a major role for strongly rough targets~\cite{Maradudin}. Nevertheless, we have also considered a non-metallic OTR radiator to ascertain the contribution of surface plasmons for the detected radiation. 
It is also important to note that, for $\beta < 0.5$ it is hardly possible to control target roughness on the scale of micro-radiator size $\beta \gamma \lambda$ for optical wavelengths under heavy ion irradiation. Already by energy deposition the ion bombardment will degrade the surface in this regime even after a short period of time. Based on the considerations above, the angular distribution, light yield and polarization of transition radiation as seen in the plane of observation from the rough target are expected to deviate from that of a smooth target. 

\section{Optical transition radiation and transverse diagnostics}
\begin{figure}[h]
\centering  
    \includegraphics[width=0.85\textwidth]{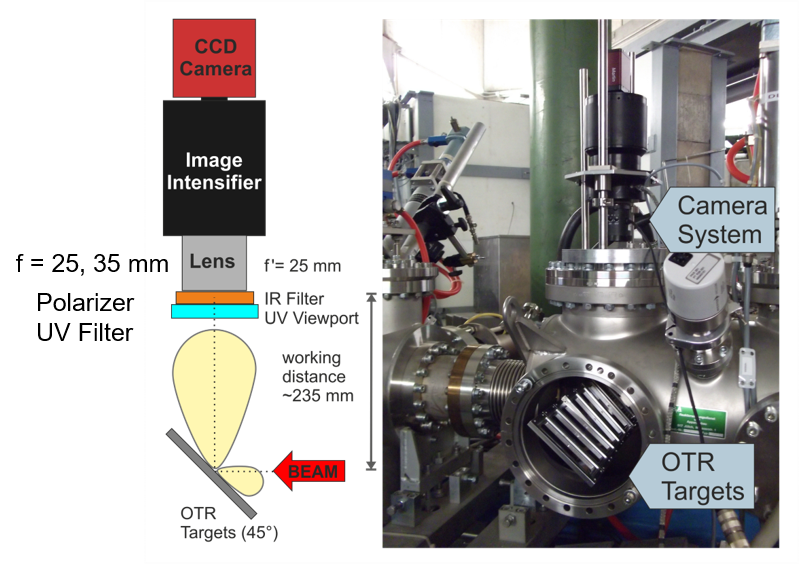}
    \caption{A set-up with a movable ladder with several OTR targets~\cite{Walasek}.}
    \label{fig:ex_schema}
\end{figure}
The experimental set-up is shown in Fig.~\ref{fig:ex_schema}. 
The metal targets used were 35~mm wide and 100~mm high. Three metals were used, aluminium, stainless steel and gold, where the former two were solid material of 2~mm thickness and the latter was in form of a 1~µm thick layer sputtered on a 2~mm thick stainless steel holder. Another target was a 5~mm thick disc (of diameter $110$~mm ) made of glassy carbon. All of them were attached to a target ladder with a stepper motor based translation stage. A rotational stage was added to the ladder during the experiments, and therefore the absolute angle of the rotational element could not be precisely calibrated. On the acquisition side, a 10 mm objective with 35 mm focal length was placed roughly $36 \pm 3$cm above the target outside the vacuum using a custom holder marked as "camera system" in Fig.~\ref{fig:ex_schema}. Thus it covers a solid angle $\Delta \Omega \approx 6\cdot 10^{-4}$ Sr. A linear polarizer was mounted directly in front of the objective. The objective is followed by an image intensifier (Proxitronic, Image intensifier BV 2582 TX-V 100 N), which was fiber coupled to a BASLER CCD camera. The pressure in the vacuum chamber was measured to be about $5 \cdot 10^{-7}$ mbar. 

\subsection{Expected number of photons in optical acceptance from a smooth target and rough targets}\label{sec:estimates}
First we estimate the total number of photons expected from backward transition radiation process in the backward half space. For a particle with given charge state $Z$ and velocity $\beta$, the total energy radiated in the frequency range of interest $\Delta \omega = \omega_2 - \omega_1$ and half unit sphere on one side of the target can be obtained be integrating the radiation intensity given in Eq.~\ref{eq:tr}.

\begin{align}
\label{eq:tr_par}
E_{rad}&=\int_{0}^{\frac{\pi}{2}}\int_{0}^{2\pi}\int_{\omega_1}^{\omega_2}{\frac{d^2I_\parallel(n,\omega)}{d\Omega d\omega} }\cdot \sin{\theta}d\theta d\phi d\omega \\ \nonumber
\end{align}
\begin{figure}[h]
\centering 
\subfigure[]{\includegraphics[width=0.45\textwidth]{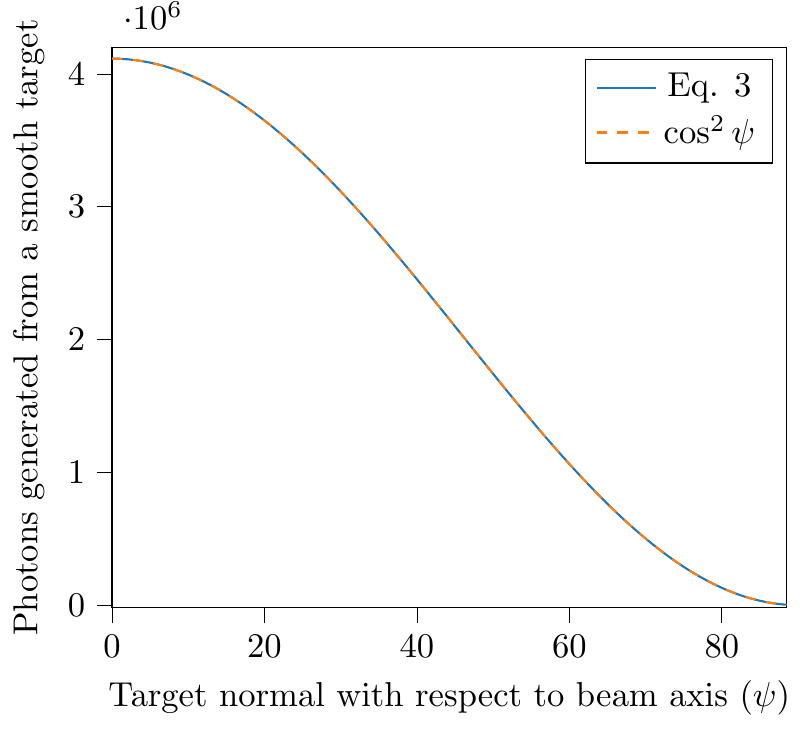}}
\subfigure[]{\includegraphics[width=0.45\textwidth]{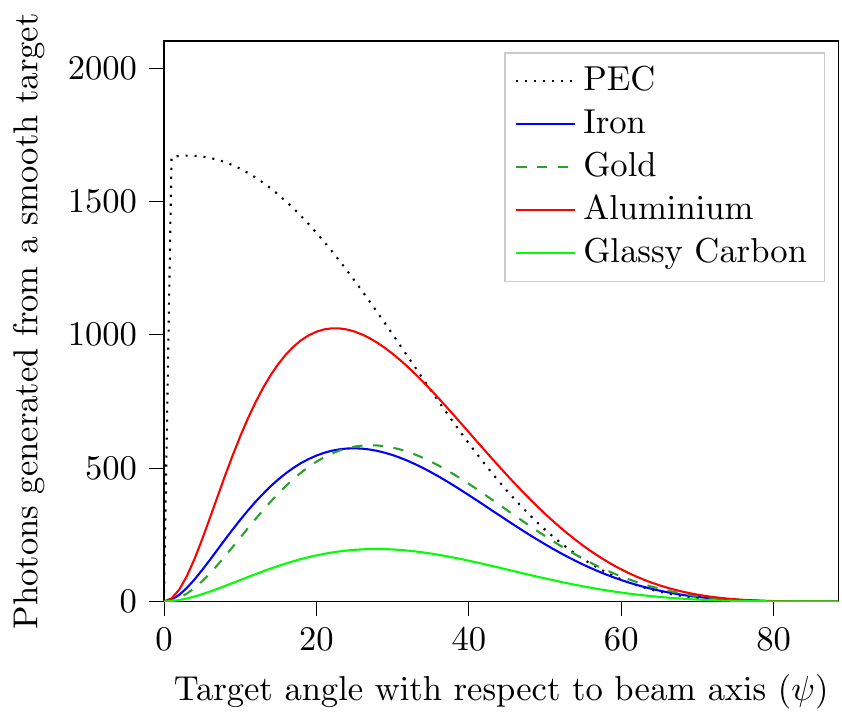}} 
    \caption{(a)Total number of photons generated in $\Delta \lambda$ = 300 nm with center at 500 nm in the half sphere as a function of $\psi$ for $5\cdot10^9$ $Ca^{10+}$ ions. (b)Expected number of photons in $6\cdot 10^{-4}$ sr from various target materials as a function of beam incidence with respect to normal $(\psi)$ at the observation angle $\theta = 90\degree -\psi$.}
    \label{fig:photon_count}
\end{figure}
Assuming a constant permittivity (see the validity of the assumption in Appendix~\ref{App:spectra}) over the frequency range and using the photon energy $E_{ph} = \hbar \omega$ the number of photons per particle is approximated by
\begin{align}
\label{eq:tr_par1}
N_{ph,t}&= \int_{0}^{\frac{\pi}{2}} \int_{0}^{2\pi} \int_{\omega_1}^{\omega_2}  \frac{\frac{d^2I_\parallel(n,\omega)}{d\Omega d\omega}}{E_{ph}} \sin{\theta} d\theta d\phi d\omega \nonumber \\[0.2cm]
&= \frac{\ln\left(\frac{\omega_2}{\omega_1}\right)}{\hbar} \int_{0}^{\frac{\pi}{2}} \int_{0}^{2\pi}  \frac{d^2I_\parallel(n,\omega)}{d\Omega d\omega} \sin{\theta} d\theta d\phi \\ 
\nonumber
\end{align}


Let us consider an example case of $Ca^{10+}$ beam with energy corresponding to $\beta = 0.11$ and an average current of 40 $\mu$A with 200 $\mu$s macropulse length which in turn corresponds to $N_{ion}=5\cdot10^9$ ions. As the spatial pulse length is much bigger than the detected wavelenghts we scale Eq.~\ref{eq:tr_par} linearly with $N_{ion}$ (incoherent sum) to find the total number of OTR photons generated by the described macropulse normally incident on a steel target as $\approx 4\cdot10^6$ photons. Due to unavailability of steel permittivity values, Iron permittivity 
is used for the calculation and it is assumed that the wavelength range of acceptance of the optical system is centered at 500 nm wavelength with $\Delta \lambda = 300 $ nm.
\begin{figure}[h]
\centering  
    \includegraphics[width=\textwidth]{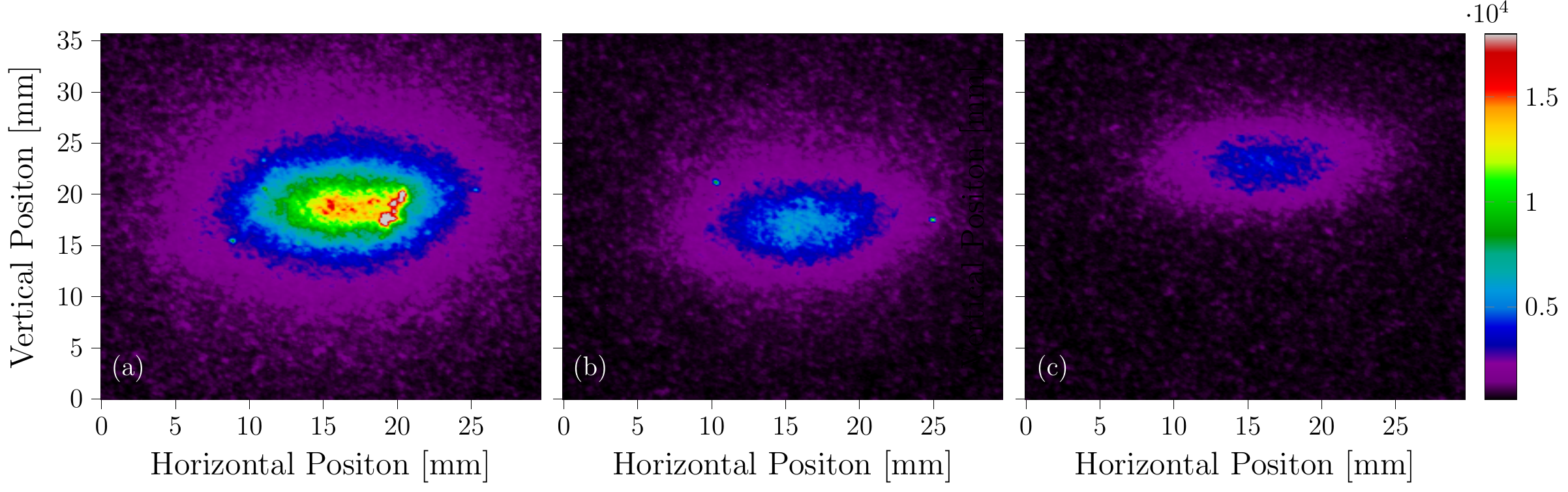}
    \caption{Transverse profile for Aluminum (top), Steel (middle) and Glassy carbon (bottom) targets for same beam conditions corresponding to $\psi= 50$ degrees for $5\cdot 10^9$ $Ca^{10+}$ ions per macropulse. The plot is an average of 250 macropulses in stable low intensity machine conditions.}
    \label{fig:profiles}
\end{figure}
Fig.~\ref{fig:photon_count}(a) shows the total photons generated as a function of beam incidence. A rough estimate for total number of photons as a function of incident ion number $N_{ions}$, energy and charge state impinging at an angle $\psi$ with respect to normal on a steel target can be given as,
\begin{equation}\label{eq:tot_photons}
    N_{ph,t} \approx 7 \cdot 10^{-4}\cdot\beta^2Z^2\cos^{2}\psi N_{ions}
\end{equation}
 The actual number of photons which will make into the optical system with a solid angle $\Delta \Omega= 0.0006$ Sr 
 as a function of beam incidence is shown in Fig.~\ref{fig:photon_count}(b).

\begin{align}
\label{eq:tr_par2}
N_{ph}&\approx \frac{N_{ion}\Delta \Omega}{\hbar} \ln\left(\frac{\omega_2}{\omega_1}\right) \frac{d^2I_\parallel(n,\omega)}{d\Omega d\omega}\\ \nonumber
\end{align}

For a discussion on the number of photons required to form a transverse profile image, see Appendix~\ref{App:HowManyPhotons}.

The conversion from well defined transition radiation generation process with a given polarization and angular distribution to an isotropic and apparently unpolarized source is governed by the surface roughness property and its interplay with the angle of incidence $\psi$. To give some qualitative arguments on the nature of this dependency, let us consider a simple rod based model for the surface roughness. 
The amount of photons generated by this rod shaped micro-radiator should have a similar scaling as smooth target, although shifted by $\alpha$ which represents the difference between $<\vec{n}_{micro,OTR}>$ and $\vec{n}_{macro}$  in the $\theta$ axis. $\alpha$ represents the slope of the rods with respect to target normal resulting in a dependence of $\cos^2(\psi+\alpha)$. 
These photons will be seen as unpolarized in observation plane ($\phi = 0$) since the "rod" normal can be aligned to any $\phi \in[-\pi/2,\pi/2]$. Generally,the dependency of light yield from an actual rough surface will differ from the qualitative picture discussed above and will depend on the shape of individual radiators as well as their distribution and separation. An accurate analysis requires a careful surface characterization and is well outside the scope of this work. We will evaluate the dependency empirically in the experimental data section. Further, one can expect that only a certain proportion of generated photons from the micro-radiators will make into the optical system due to scattering due to surrounding structures and an additional $\sin{\psi}$ dependence can be expected. This is similar to the optical system property "Etendue" arising due to relative angle between the source and detector. Apart from the aforementioned "geometrical" factors; another potential mechanism for dependence of the light yield as a function of irradiation angle is the diffraction radiation and its role can become increasingly dominant as $\psi \to \pi/2$ as mentioned in~\cite{Arutyunyan}. If the charge travels close to the rough surface without actually hitting it, a significant enhancement in the observed radiation for grazing angles is possible. In case of periodic structures on the surface, a peak in spectra corresponding to periodicity due to constructive interference of the radiation can also be expected.


\begin{figure}[h]
\centering  
    \includegraphics[width=0.45\textwidth]{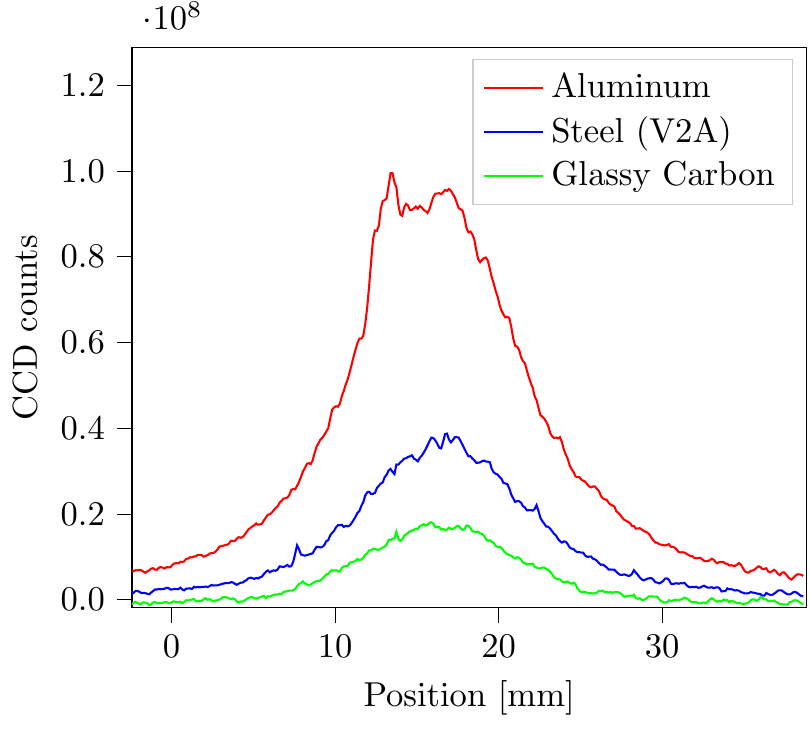}
    \caption{Beam profile in horizontal plane for Aluminum, Steel and Glassy carbon targets for $\psi= 50$ degrees for $5\cdot 10^9$ $Ca^{10+}$ ions.}
    \label{fig:photon_count_hor}
\end{figure}
Figure~\ref{fig:profiles} shows the two dimensional beam images from the transition radiation for three target materials, Aluminum, Steel (V2A) and Glassy carbon under the same beam and acquisition conditions. $5\cdot 10^9$ $Ca^{10+}$ ions was irradiated on the targets at an irradiation angle of $\psi = 40$ degrees. An average over $250$ images was performed to obtain the two dimensional images where the colorbars are fixed to same values for all of the three images. The photon count scaling trend for different target materials is similar to Fig.~\ref{fig:photon_count} (but not the same) and is also visible in the profile heights of the horizontal projection plotted in Fig.~\ref{fig:photon_count_hor}. The discrepancy is mainly for Steel target where we have used permittivity values for Iron. The measured profile width is roughly the same for all targets. We also see, that there is a saturated region on the Aluminum target at the co-ordinates (20 mm, 20mm) (Fig.~\ref{fig:profiles}), and could be related to macroscopic surface structure on the target. Similar less pronounced spots are seen at (10mm, 22mm) and (25mm, 17mm) on the steel target. Another observation is that the background is proportional to the peak height, which hints that the background might be primarily composed to scattered OTR photons.

The CCD count itself is a strongly non-linear function of intensifier voltage and cannot be directly correlated with the number of photons (discussed in Appendix~\ref{App:OpticalSystem}). However, for multiple measurements under the same beam conditions, the CCD count fluctuation can provide an estimate of the average number of photons generated. This would however be only true for lower photon yields where the CCD count fluctuation is dominated by the statistical fluctuation in the photon counts. This is discussed further in the Appendix~\ref{App:ShotNoise}. 


\subsection{CCD counts vs beam current}
\begin{figure}[h]
\centering  
    \includegraphics[width=0.65\textwidth]{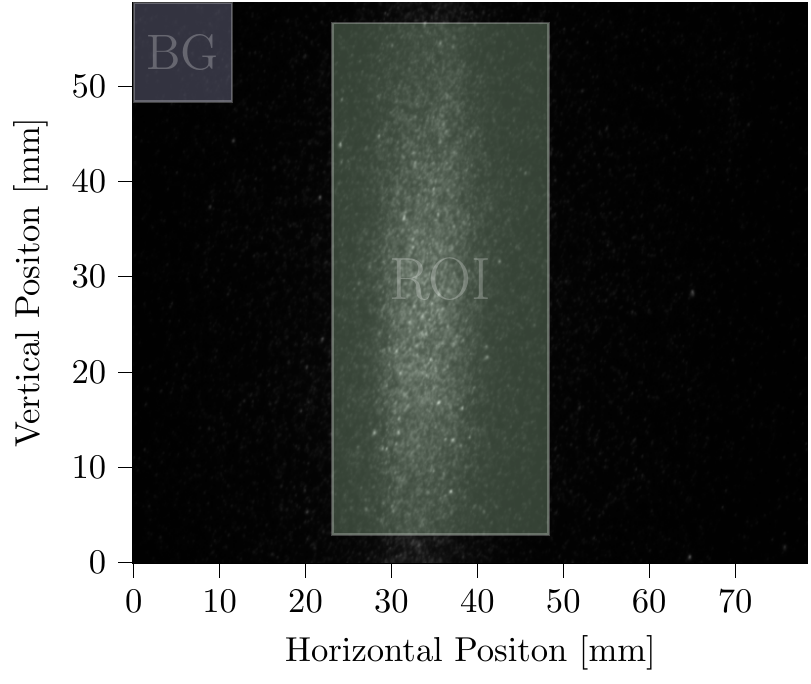}
    \caption{Individual image taken with a 80 $\mu$s gate for 200 $\mu$A $Bi^{26+}$ beam current. The rectangle marked on top left is used to subtract the background counts per pixel.}
    \label{fig:BG_ROI}
\end{figure}

\begin{figure}[h]
\centering
\subfigure[]{\includegraphics[width=0.45\textwidth]{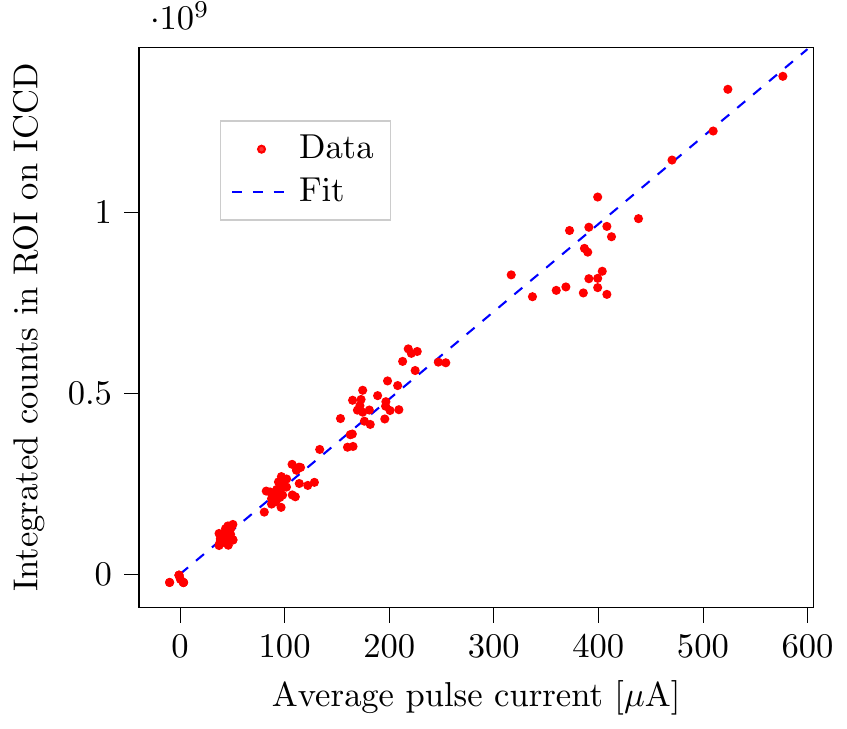}} 
\subfigure[]{\includegraphics[width=0.45\textwidth]{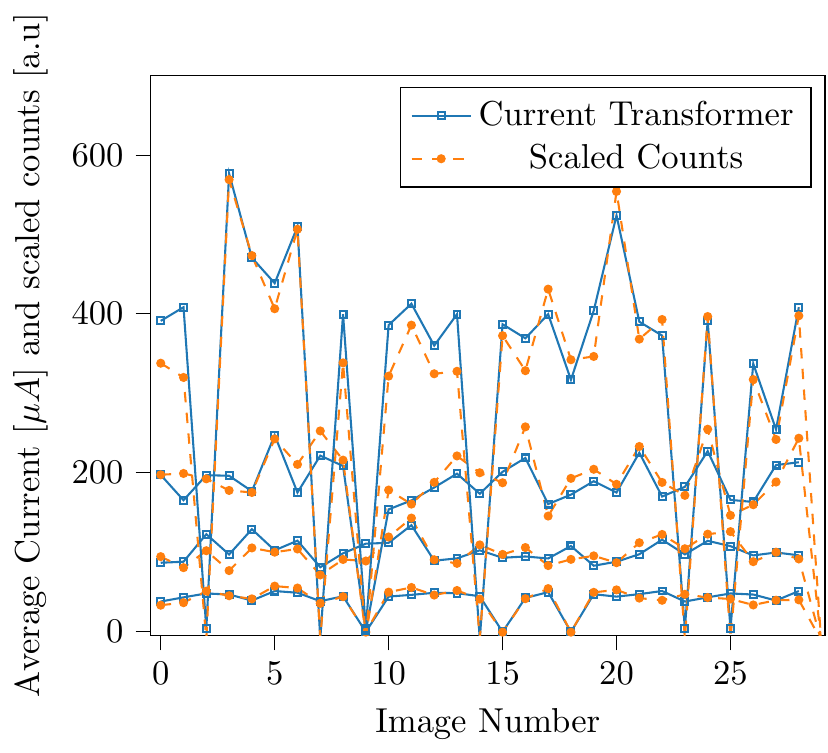}} 
\caption{(a) Macropulse average current vs integrated count per image in the region of interest. A linear behavior from $40 \mu A$ to $600 \mu A$ (b) Current transformer reading vs integrated counts on consecutive images for the four current settings.}
\label{fig:cur}
\end{figure}

A systematic study to observe the relation between average beam current against number of counts for the same intensifier and CCD settings. This study was incidentally performed with a Gold target ($Au$) with Bismuth $Bi^{26+}$ beam. Figure~\ref{fig:BG_ROI} shows a single image for $200$ $\mu A$ beam current at $\psi = 70$ deg. The background counts are subtracted by using the average pixel value from outside the beam irradiation. This background region (BG) and region of interest (ROI) ares annotated. 
   An average beam current to CCD count dependence is shown in Fig.~\ref{fig:cur} (a). Each data point corresponds to a current transformer reading and the corresponding image counts in the region of interest (ROI) for the given macropulse. There is clear linear dependence between beam current and CCD counts. Figure ~\ref{fig:cur} (b) shows temporal correlation of the beam transformer current and scaled CCD counts for ROI in the image for the same data as in Fig~\ref{fig:cur}. The conversion from photons entering the intensifier input to CCD counts is discussed in the Appendix.

\subsection{Polarization study}
The dominant component of transition radiation due to a smooth target for low beta beams is linearly polarized in the plane of radiation. This is one of the signatures of transition radiation in comparison to other photon inducing charged particle interactions like beam induced fluorescence (BIF) or material luminescence.
\begin{figure}[h]
\centering
\includegraphics[width=\textwidth]{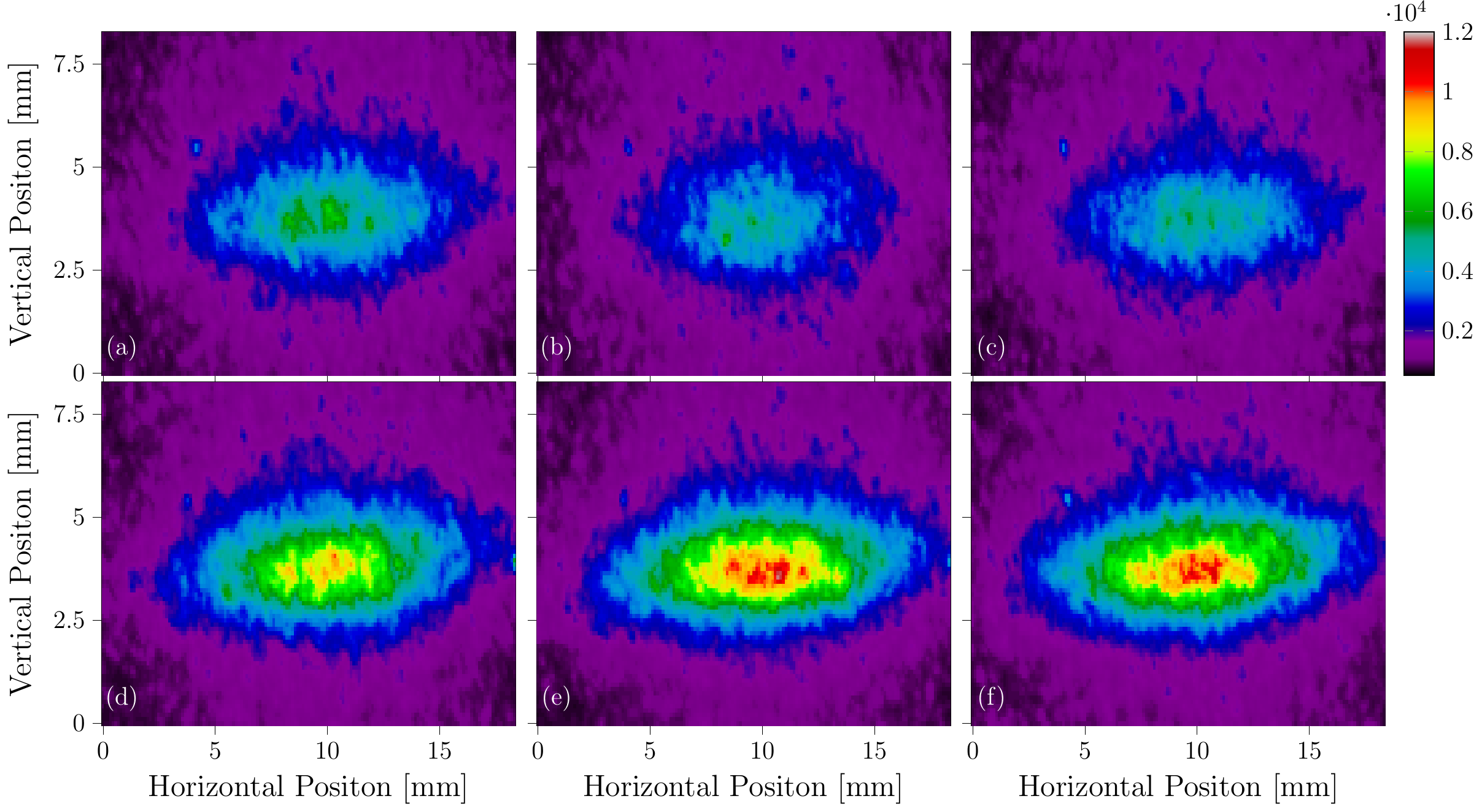}
\caption{Image with a steel (V2A) target at 20 degrees with respect to normal at the polarization angle of (a) 0 (b) 30 (c) 60 (d) 90 (e) 120 and (c)150 deg.}
\label{fig:v2a80}
\end{figure}
\begin{figure}[h]
\centering  
    \includegraphics[width=\textwidth]{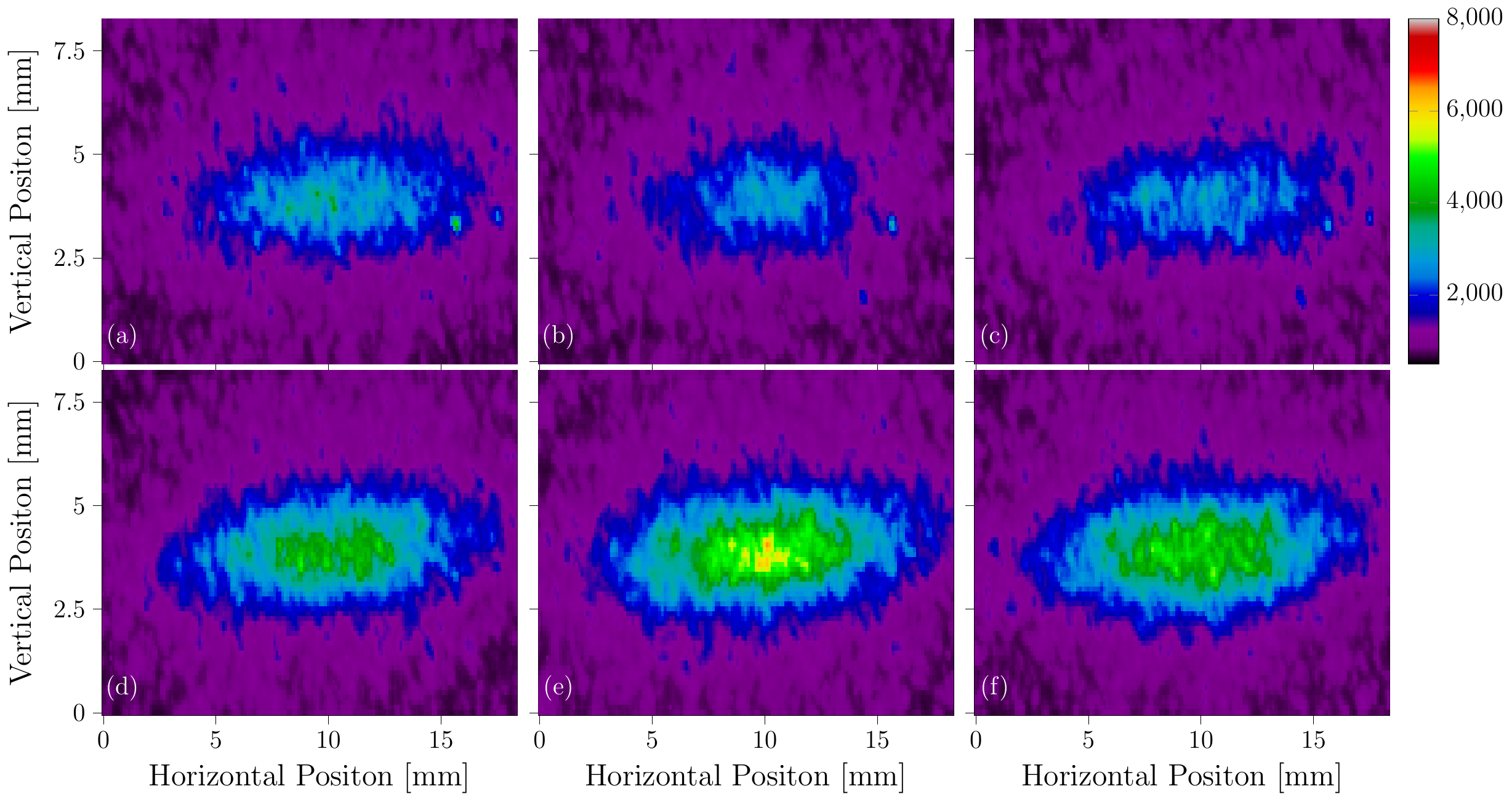}
    \caption{Image with a Glassy carbon target at 20 degrees with respect to normal at the polarization angle of (a) 0, (b) 30, (c) 60 (d) 90 (e) 120 and (f) 150 deg.}
    \label{fig:gc1}
\end{figure}
\begin{figure}[h]
\centering 
\subfigure{\includegraphics[width=0.48\textwidth]{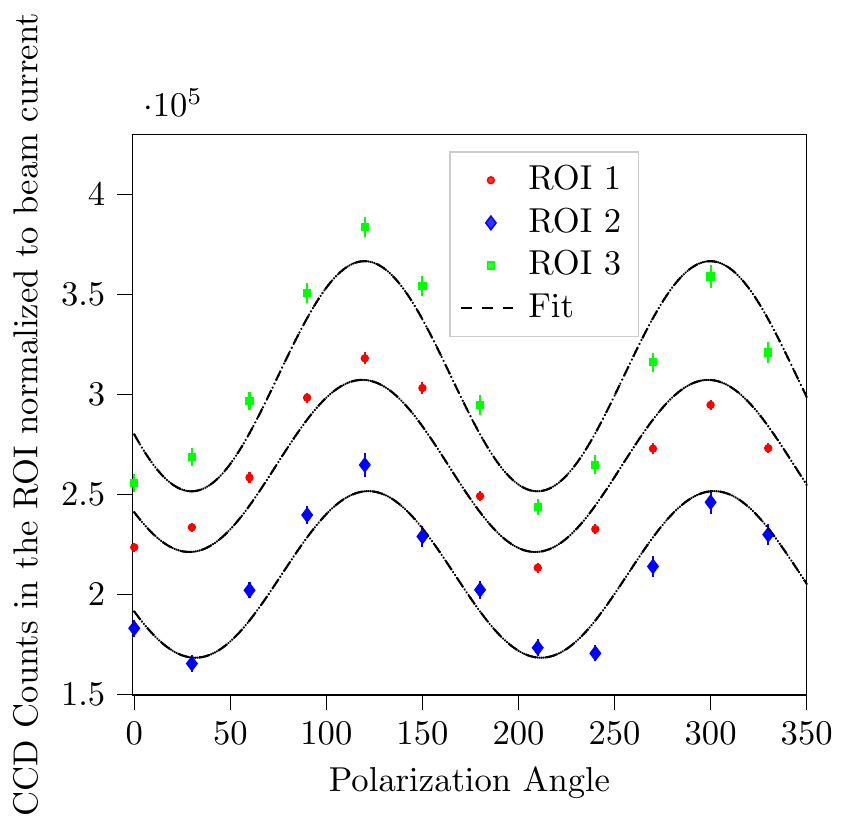}}
\subfigure{\includegraphics[width=0.48\textwidth]{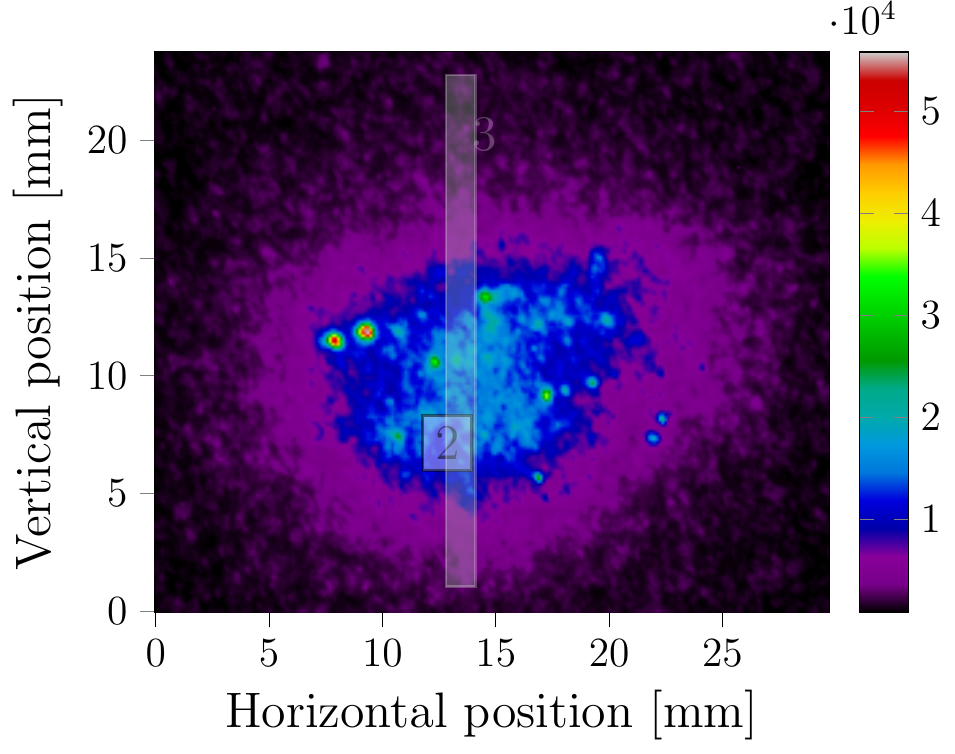}} 
 \caption{OTR light output from an Aluminum target as a function of polarizer angle. 120 and 300 degrees on the polarizer corresponds to plane of incidence. The different regions of interest for calculated counts are shown in the left figure. ROI 1 corresponds to the shown full image while ROI 2 and ROI 3 are smaller sections which are marked.}
    \label{fig:alu_pol}
\end{figure}
We measured images for different polarizer settings with respect to plane of observation for a macropulse containing $5\cdot10^9$ $Ca^{10+}$ ion beam with velocity $\beta = 0.11$ irradiated on three target materials (Aluminum, Steel and Glassy Carbon). One image was captured per macropulse. The MCP gain was set to 1.726 V.
Figures~\ref{fig:v2a80} and~\ref{fig:gc1} shows 2D transverse profiles for $\psi = 20$ deg at six consecutive polarizer angles from a steel and glassy carbon target respectively. The polarizer angle $= 120\degree$ is when the polarizer axis is parallel to plane to observation. These images were obtained by averaging 250 images. The number of CCD counts are generally factor $\approx 2$ higher for Steel target in comparison to Glassy carbon target. There is a component which is polarized in the plane of observation for both steel and glassy carbon target. 
Observing linearly polarized radiation from both metallic and dielectric targets can be considered as the evidence of OTR being a dominant radiation mechanism in vacuum for the given beam conditions. It is also worth noticing that, in the case when polarizer angle is perpendicular to the plane of observation $( 30 \degree)$, there is still a significant amount of radiation left and the image corresponds well to the beam profiles with other polarizer settings.  
This can be understood as a consequence of the aforementioned surface roughness where the planes of radiation of micro-radiators do generally not coincide with the one of the macroscopic target. Therefore micro-radiators will effectively contribute to the photon yield which is unpolarized with respect to the plane of observation.

Figure~\ref{fig:alu_pol}(a) shows the average CCD counts for 250 images as a function of polarization for an irradiation angle of $\psi = 30\degree$ on an Aluminum target. The polarizer angle was incremented in $30\degree$ steps at for each set of measurements and covers the full 360 degrees. For $N$ image measurements under the same beam conditions, the error bar on the mean value of CCD counts is gives by $\delta_{counts}=\sigma_{counts}/\sqrt{N}$.  The polarization angles 120\degree and 300\degree correspond to the plane of observation. The counts are accumulated for three different regions of interest (ROIs). ROI 1 corresponds to the full image shown in  Fig.~\ref{fig:alu_pol}(b). The counts from ROI 1 are reduced by factor 10 for better visualization. ROI 2 and 3 are marked in the figure. They were chosen to avoid the "hotspots" on the image. These hotspots occur the at same location on the target irrespective of beam movement, and therefore we suspect that they occur due to surface non-uniformity forming some sort of a photon concentrator. There is no dependence on the relative contributions of polarized to unpolarized light with the choice of ROI. However, it is also worth noticing that in this specific case of Aluminum for $\psi = 30\degree$, there is roughly a factor $\approx 8$ more unpolarized radiation (since half of unpolarized photons are blocked by the polarizer). The number of photons from Aluminum target for steep angles $\psi<30\degree$ was saturating the CCD images since most of the photons were hitting the few pixels exposed in the vertical plane. Further, the Aluminum target has these hotspots exactly on the location where beam was hitting as shown in Fig.~\ref{fig:profiles}. We have therefore not considered the data from the Aluminum target for target rotation and angular distribution studies discussed in the following section.

\subsection{Light yield and angular distribution }
\begin{figure}[h]
\centering
\includegraphics[width=\textwidth]{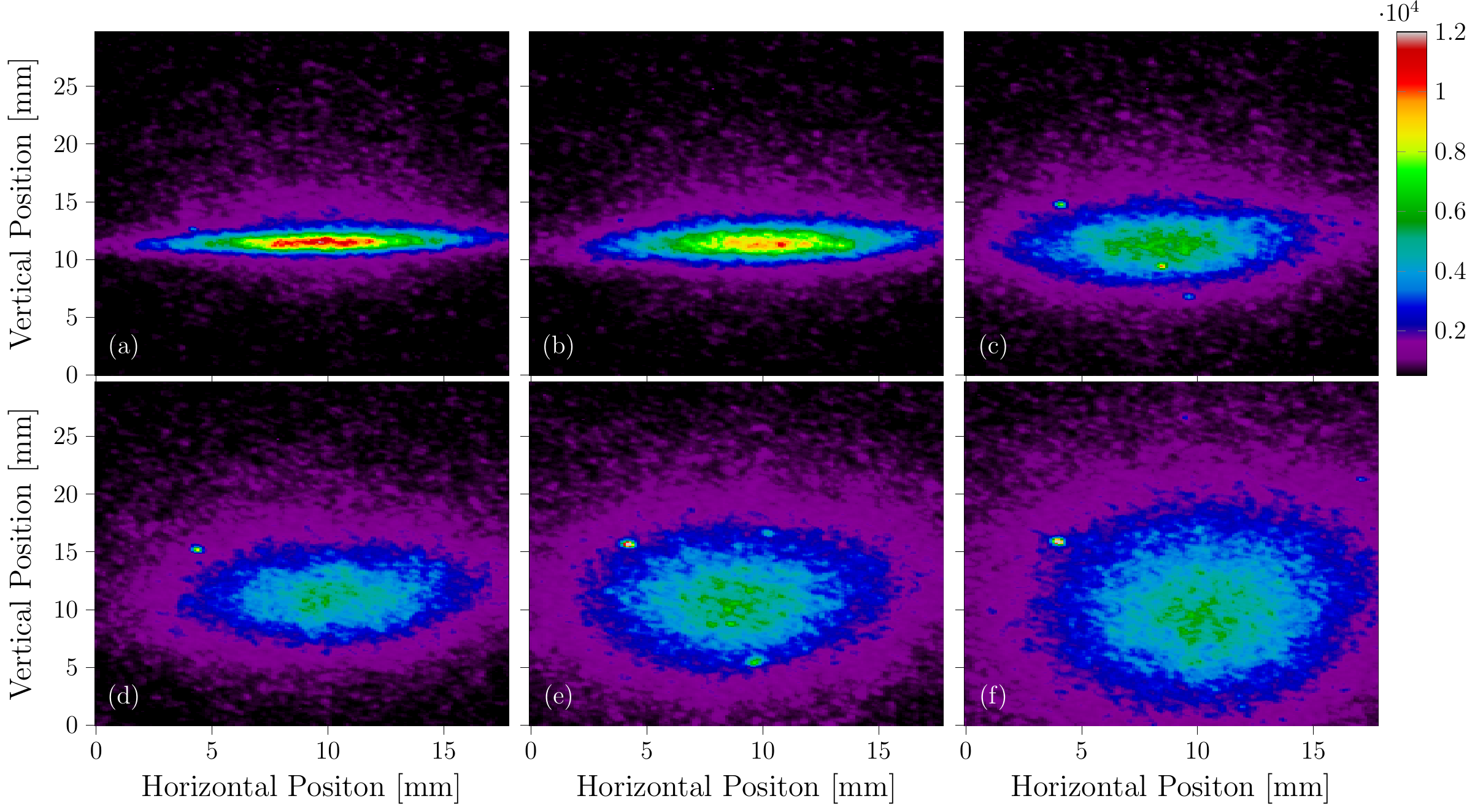}
\caption{Image with a steel (V2A) target at (a) 10 (b) 20  (c) 40 and (d) 50 (e) 60 and (f) 70 degrees. with respect to normal at the polarization angle of 120 deg.}
\label{fig:v2a_ang}
\end{figure}

Figures~\ref{fig:v2a80} and~\ref{fig:gc1} showed the polarization for a fixed irradiation angle $\psi=20\degree$. In the next step, we measured the radiation from several target angles in the range $\psi=10-70\degree$ along with a systematic variation in the polarizer settings covering 180 degree rotation in six steps. For a fixed polarizer angle, Fig.~\ref{fig:v2a_ang} shows the beam image on a steel target as a function of $\psi$. The vertical projection of the beam increases expectedly when the irradiation angle becomes larger moving towards the grazing angle.
Since the target was rotated, the observation angle $\theta$ is correlated with irradiation angle $\theta = 90 - \psi$ and thus not an independent parameter. 

\begin{figure}[h]
\centering 
\subfigure{    \includegraphics[width=0.45\textwidth]{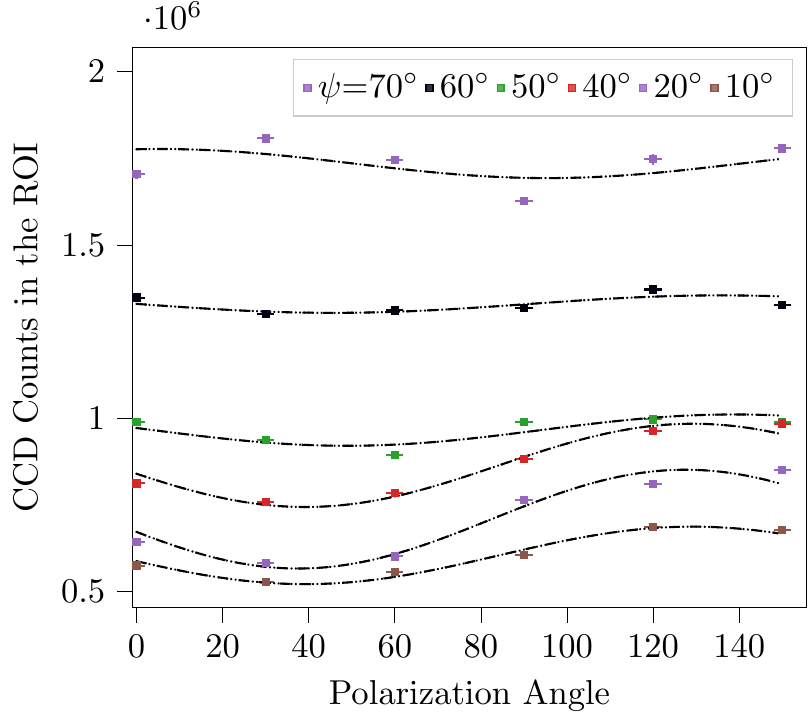}} 
\subfigure{\includegraphics[width=0.45\textwidth]{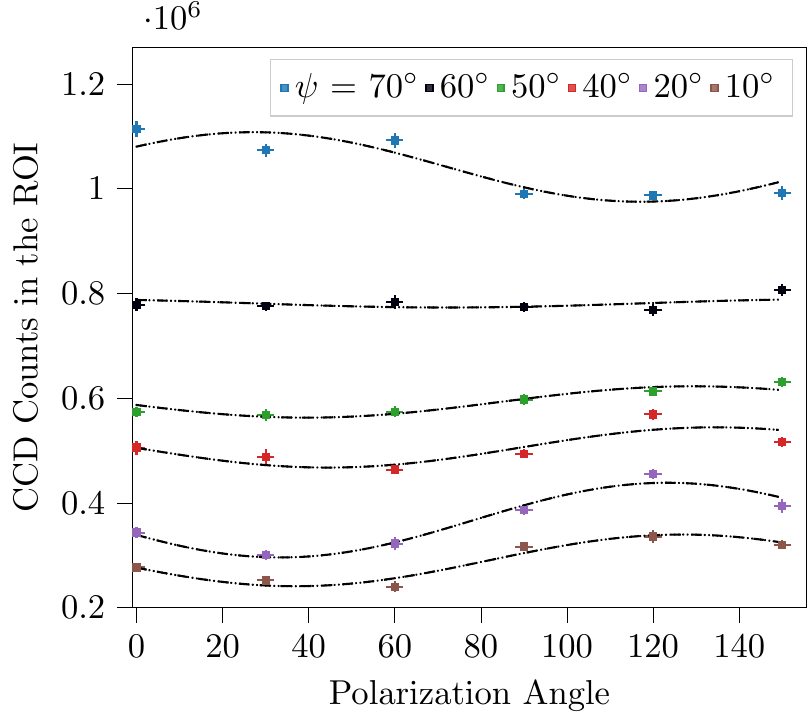}} 
\caption{CCD counts as a function of incidence angle and polarization angle for steel target (left) and Glassy carbon target (right).}
\label{fig:Pol_steel_Steel_GC}
\end{figure}

Figure~\ref{fig:Pol_steel_Steel_GC} shows the fit separating polarized CCD counts and unpolarized CCD counts vs polarization angle for several irradiation angles for Steel and glassy carbon targets. In spite of the large error bars, the polarized components could be fitted with the typical polarization curve and the peak at most irradiation angles is seen at the polarizer angle of 120\degree when the polarizer axis coincides with the plane of observation. A peculiar observation is that, at $\psi > 60\degree$, we see a shift in polarization direction and an emergence of polarized photons perpendicular to the plane of observation. This observation suggests that the mean of microscopic normal distribution $<\vec{n}_{micro,OTR}>$ varies as function of irradiation angle and becomes perpendicular to the plane of observation for ~$\psi > 60\degree$ resulting into a change of net polarization.

\begin{figure}[h]
\centering 
\subfigure[]{\includegraphics[width=0.45\textwidth]{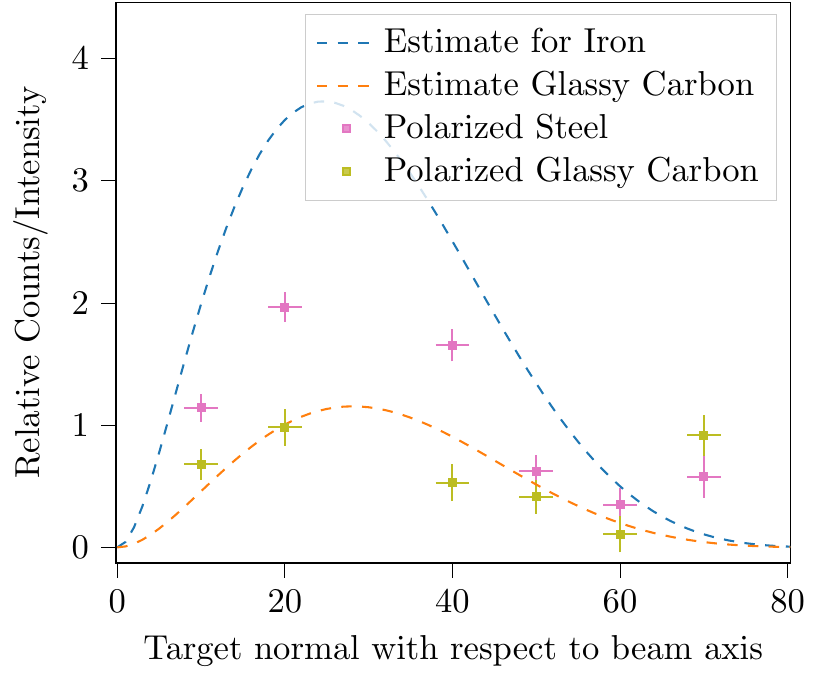}} 
\subfigure[]{\includegraphics[width=0.45\textwidth]{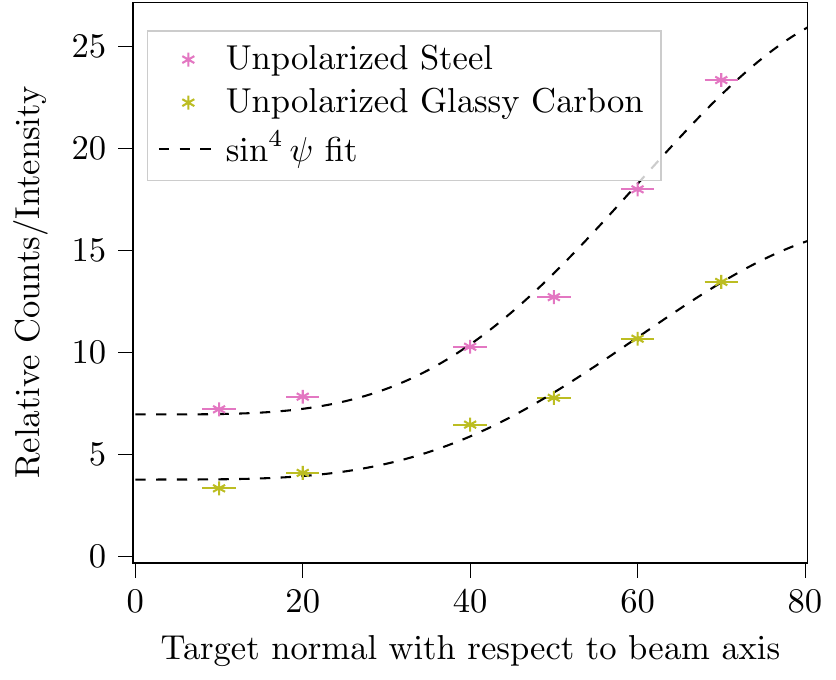}} 
\caption{(a) Detected polarized photons from steel (V2A) and Glassy carbon target as a function of polarizer and target angle. (b) Unpolarized photons from steel (V2A) and Glassy carbon target as a function of polarizer and target angle.}
    \label{fig:va2_alu}
\end{figure}

In Fig.~\ref{fig:va2_alu}, we have summarized the linearly polarized and unpolarized radiation with respect to plane of observation  as a function of irradiation angles for steel and Glassy Carbon from the data shown in Fig.~\ref{fig:Pol_steel_Steel_GC}. The linearly polarized light in observation plane is compared with the scaled theoretical estimate of photons shown in ~\ref{fig:photon_count} as a function of $\psi$. The plots for the estimates are normalized to match the measured sample for glassy carbon at $\psi = 10$. 
The polarized light is seen when the charges hit the surfaces whose normal vectors $\vec{n}_{micro,OTR}$ lie in the plane of observation as expected from the macro normal $\vec{n}_{macro}$. There is a rather good agreement for Glassy carbon data. For the Steel target, the trend against $\psi$ is correctly reproduced but the absolute estimate value has a notable discrepancy at $\psi = 20 \degree$. 
As mentioned previously the permittivity values for Iron were used to estimate the theoretical light yield over irradiation angle to be compared to the Steel target.
Even for Iron the permittivity values given in literature have a rather large variation~\cite{Ordal,Christy,Querry}. As already discussed earlier, above $\psi>60$, there are more micro-radiator surfaces perpendicular to the plane of observation than the ones parallel to it giving a net linearly polarized contribution in that direction.

The unpolarized component of radiation shown in Fig.~\ref{fig:Pol_steel_Steel_GC}(b) increases with $\psi$ and could be fit satisfactorily with the lowest order dependence of $\sin^4{\psi}$. 
Similar observations were also made for Gold and Aluminum targets.  The measured dependency is atleast an order higher than qualitatively argued i.e. $\cos^2({\psi+\alpha})\sin{\psi}$ discussed in section~\ref{sec:estimates}. The increase in photon counts at shallower angles (as $\psi \to \pi/2$) can be a cumulative effect of many distinct effects.
\begin{itemize}
    \item The $\cos^2({\psi+\alpha})$ dependence of generation process with the irradiation angle.
    \item Lower scattering for the photons generated by micro-radiators at grazing angles because the probability of shielding of photons by the generating structure as well as rough neighborhood is reduced.
    \item An increase in the diffraction radiation is likely at grazing angles. This is supported by another experimental observation, where we have observed a large amount of radiation when the edge of targets were irradiated. With diffraction playing a large role for shallow angle irradiation, a bound dependency like $\sin^4{\psi}$ will most likely break down.
    \item  Significant radiation at grazing angles is directed towards the target surface which can be specularly reflected or scattered towards the detector, resulting is further increase in light yield.
\end{itemize} 
Given the data at hand, it is difficult to distinguish the dominant component which results into the increased unpolarized radiation among all the effects discussed above. Further, we have observed that under heavy ion irradiation, the polarized components of radiation reduces further while unpolarized component is unaffected. Faster surface deformation due to heavy ion irradiation might reduce the polarized contributions arising from $\vec{n}_{macro}$ part of the target.

However the aforementioned observations support our working hypothesis; i.e. a vast majority of the detected radiation on the CCD is indeed the optical transition or diffraction radiation due to the target surface roughness. The hypothesis is further strengthened by the measurements from the Glassy Carbon which is a radiation-hard dielectric material. The material properties of Glassy carbon allows no other generation mechanism like surface plasmons or metallic oxide fluorescence. The yield for polarized, unpolarized and background light scales expectedly with the permittivities of different target materials and hints strongly that all the radiation (including background) is primarily transition or diffraction radiation. From the transverse profile measurement perspective, there are only two relevant questions; 1) Is the observed un-polarized radiation a result of transition or diffraction radiation processes? 2) Does it represent the exact transverse profile of the beam. Based on the data available, the answer to both the questions is affirmative.

\subsection{Comparison of OTR profile with SEM-Grid}
\begin{figure}[h]
\centering 
\subfigure{    \includegraphics[width=0.45\textwidth]{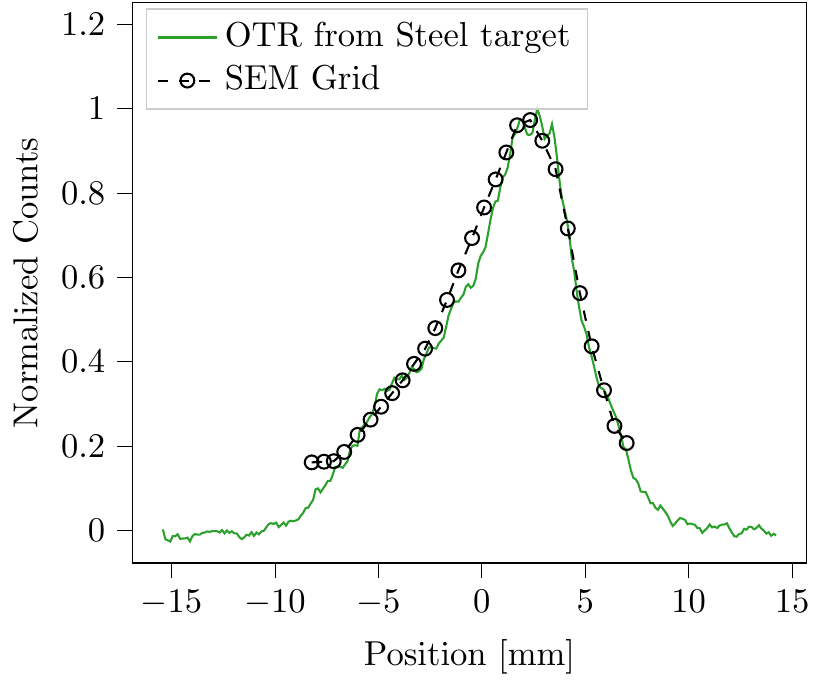}} 
\subfigure{\includegraphics[width=0.45\textwidth]{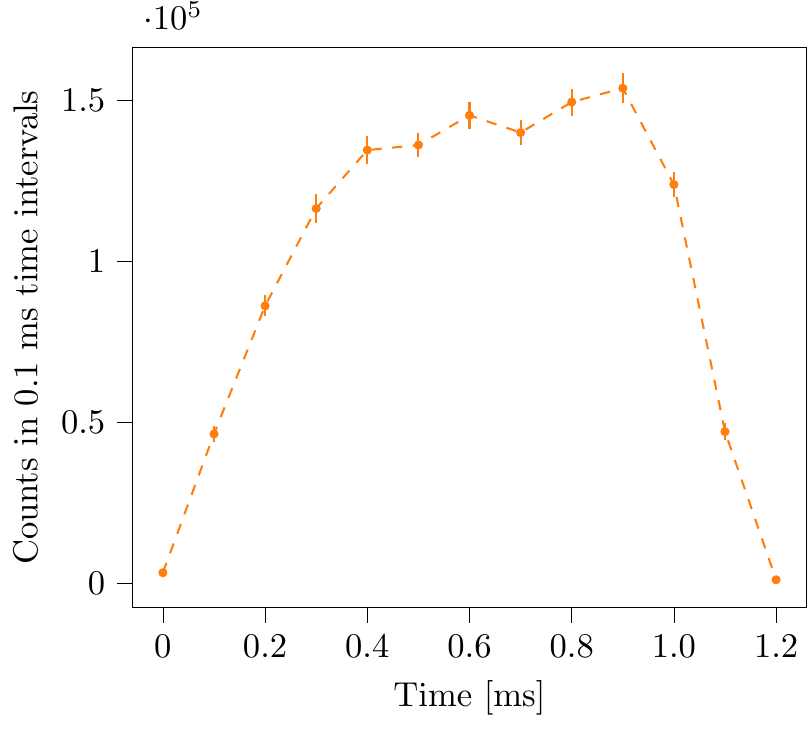}} 

    \caption{ (a) Comparison of a horizontal profile from secondary electron emission grid (SEM-Grid) and OTR light for a $Ar^{18+}$ beam with 95 $\mu$A and 100 $\mu$s pulse length. (b) The mean count for $Fe^{25+}$ beam with 20 $\mu$A average current in a 1 ms macropulse for 100 $\mu$s MCP gate width with a varing delay marked on the x-axis. The error bars are calculated from count fluctuations over 100 macropulses.}
    \label{fig:OTR_SEM}
\end{figure}
A comparison between secondary electron emission monitor grid (SEM-Grid) and OTR image which are longitudinally separated by ~1 m is shown in Fig.~\ref{fig:OTR_SEM} (a). This measurement was performed with 8.6 MeV/u 95 $\mu$A $Ar^{18+}$ beam irradiated on a steel target. The SEM-Grid wires are 2.1 mm apart and and interpolation between the wires is performed by the SEM grid software. Generally a good agreement between OTR images and SEM-Grid profiles is seen.
\subsection{Counts a function of MCP gate delay}
As part of the initial studies to rule out fluorescence and any other slow processes, MCP gate was set to a fixed width (100 $\mu s$) and the measurement was triggered at different delays with respect to beam macropulse arrival.  This measurement was performed with 11.4 MeV/u 20 $\mu$A $Fe^{25+}$ beam with a macropulse length of 1 ms. The coarsely sampled pulse shape was reconstructed with this measurement in 13 delay steps as shown in ~\ref{fig:OTR_SEM} (b).

\subsection{Target heating and thermal photons}

\begin{figure}[h]
\centering  
    \includegraphics[width=0.45\textwidth]{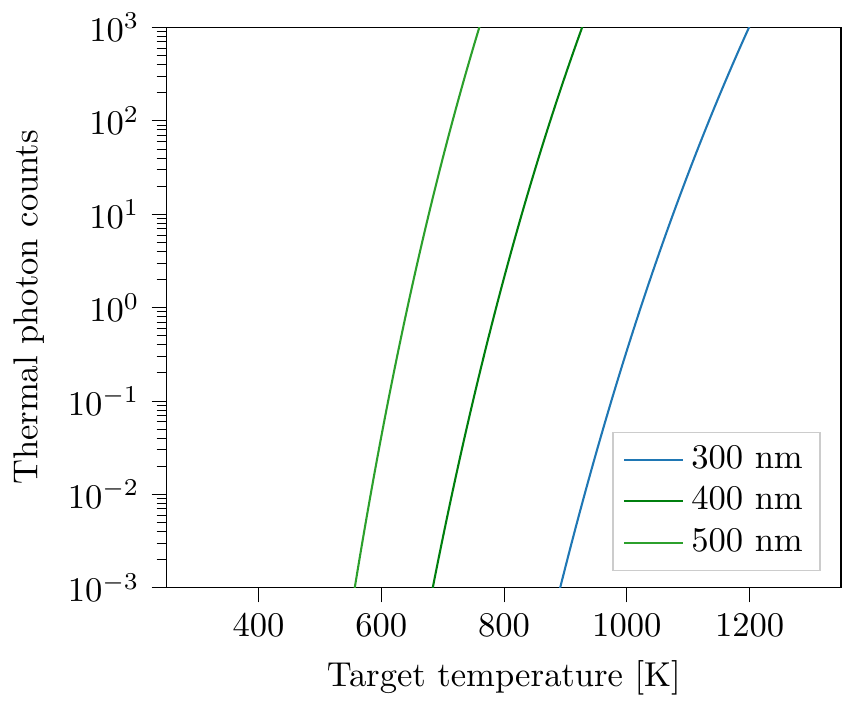}
    \caption{Thermal Photons.}
    \label{fig:thermal_photons}
\end{figure}
The number of OTR photons per image are in the order of 20-1000 for low intensity beams discussed in this report. Therefore profile imaging using OTR is sensitive to any thermal photons resulting from target heating. An estimate of thermal photons per unit frequency $f$ and solid angle $\Omega$ as a function of target temperature can be obtained by Planck's law, 
\begin{equation}\label{eq:thermal}
\frac{dI_T(n,f)}{d\Omega df}= \frac{2hf^3}{ c^2}\cdot \frac{1}{e^{hf/K_BT}-1}
\end{equation}
Fig.~\ref{fig:thermal_photons} shows the dependence of the number of photons generated as a function of target temperature for radiation centered in different parts of the spectra within $200$ nm spectral width. One can see a threshold like behavior; i.e. when the temperature of target crosses a certain threshold, thermal photons with longest wavelengths allowed by the optical system  will start to interfere with the measured profile image. As can be seen that already at ~700 K temperature, photons centered at 500 nm photons can overwhelm the OTR based profile image.
\begin{figure}[h]
\centering 
\subfigure{\includegraphics[width=0.45\textwidth]{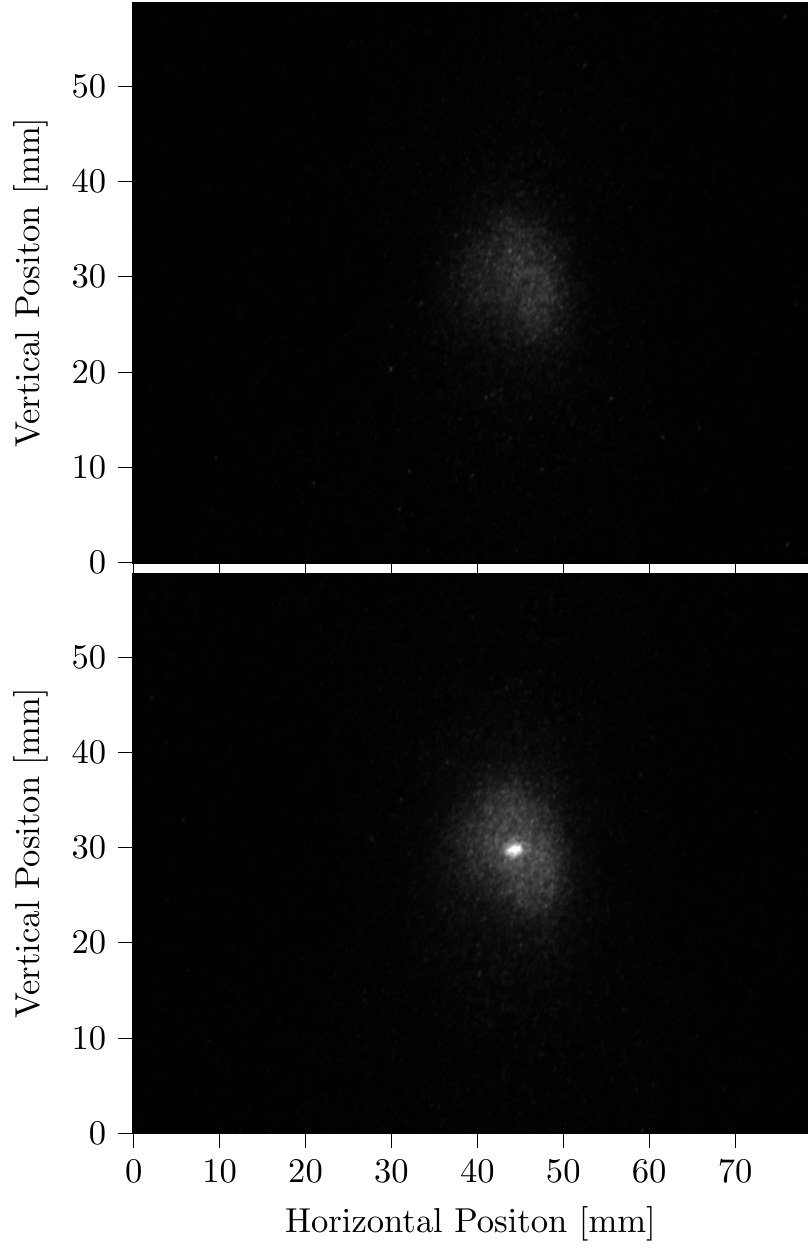}} 
\subfigure{\includegraphics[width=0.45\textwidth]{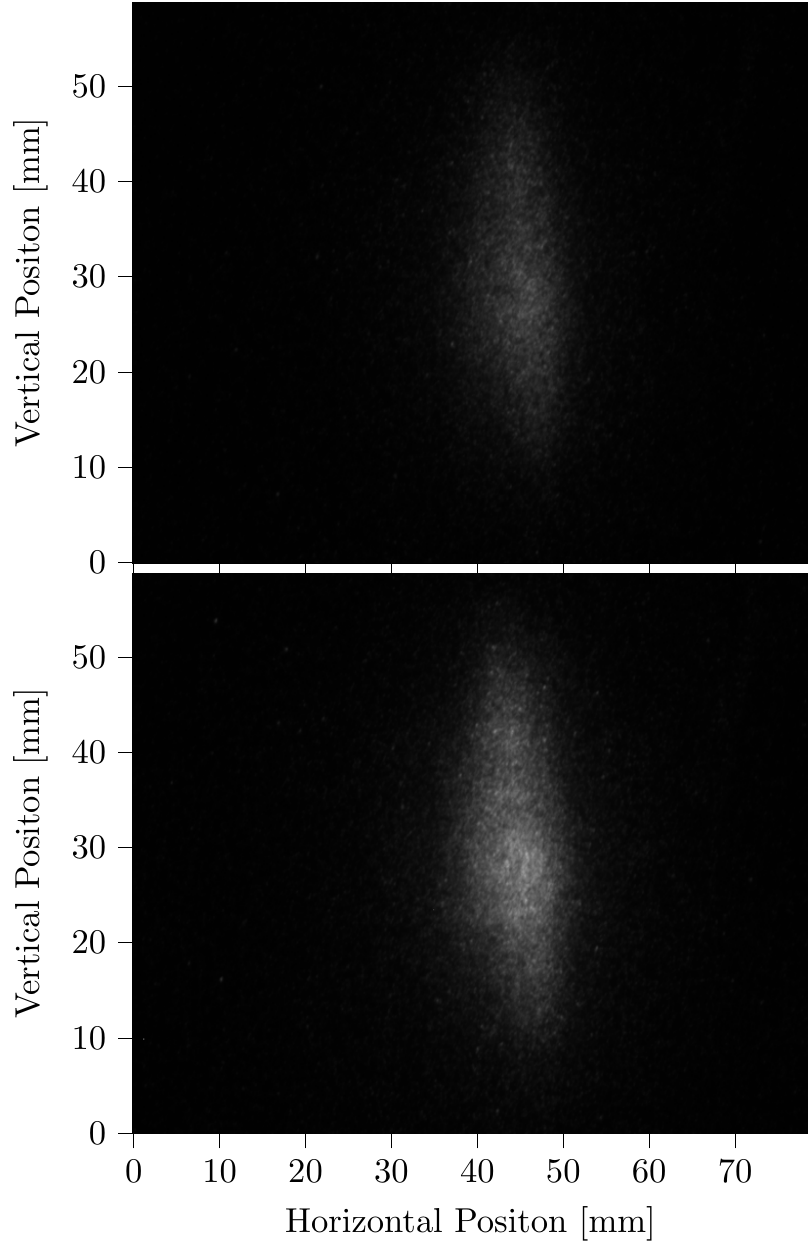}} 
    \caption{Image with a 70 us gate (top) and 100 us (bottom) from the beginning of the macropulse for 400 uA current for target incidence $\psi= 25$ deg.Image with a 70 us gate (top) and 100 us (bottom) from the beginning of the macropulse for 400 uA current for target incidence of $\psi= 65$ deg.}
    \label{fig:thermal_photon1}
\end{figure}
This behavior was observed with a 400 uA current $Bi^{26+}$ beam as shown in Fig~\ref{fig:thermal_photon1}(a). There was no optical filter applied for this specific measurement. Using a ICCD gating period of 70 us, the beam profile is not distorted while with a gate of 100 us, the central part of the image is dominated by thermal photons. This concludes that the center of the target with beam crosses the temperature threshold in about ~70 us from the start the of macropulse. Target heating can be counter-acted by depositing the beam energy of a larger surface area of the target as shown in Fig.~\ref{fig:thermal_photon1}(b), where under exactly the same beam conditions as Fig.~\ref{fig:thermal_photon1}(a), no heating is observed. 
Since a larger area of the target was under irradiation, the target did not cross the temperature threshold to generate enough thermal photons. Thermal photons can be reduced by utilizing either shorter macropulses, shallower target angles, infrared filters or active target cooling. As shown in Fig.~\ref{fig:thermal_photons}, application of optical filters which cut out photons above 300 nm can allow the target temperature to increase upto 1100-1200K without any significant thermal photon disturbance. For most practical measurements in our current range of upto few mA, target heating should not be an issue if marcopulse length is controlled or an optical filter is applied. However, target heating always needs to be considered in an OTR based diagnostics for low energy hadron beams.
 

\section{Summary and applications}

We have shown that the OTR for ion beams provides enough photons for the measurement of a beam profile covering the typical range of intensities in accelerator operations. The light yield and polarization differs significantly between an ideal smooth target and the rough targets as observed experimentally. Rough targets has enhanced radiation at grazing angles, which can lower the intensity and energy thresholds for usability of the OTR process in hadron beam diagnostics. It also opens up possibilities for optimization of camera angles for beam imaging, non-destructive profile monitoring and energy deposition on a larger surface area to avoid target heating. There are few immediate applications, the first is the construction of a SEM-Grid like profiler that will allow SEM-Grid like diagnostics especially since the light emissions from the edges of the wires (grazing incidence) can be quite high. It could be advantageous for transversally small high intensity beams, since construction of SEM-Grids is particularly challenging in those cases. The second use case is the modeling of non destructive devices like ionization profile monitors (IPM) and beam induced fluorescence (BIF) monitors under high beam intensity. The IPM and BIF monitors are known to be affected by direct beam fields (space charge) and OTR can provide an in-situ non space charge affected profile for correction and modeling. Eventually, with the usage of very thin foils ($< 0.5$ $\mu$m) such that beam traverses without significant energy deposition, transverse profile measurements can be combined with bunch length measurements using recently demonstrated GHz transition radiation based bunch length monitor~\cite{Singh} to obtain a fast, non-destructive and compact full 6D emittance measurement set-up. Finally, OTR with thin radiation hard materials like Silicon carbide (SiC), Glassy Carbon, Carbon stripper foils (already used at GSI~\cite{Barth}), Zinc Oxide (ZnO) in high energy beam transport opens up a possibility for an almost non-destructive profile monitoring especially for fully stripped particles. The light yield after the synchrotrons should be high enough to be observed on a normal CCD directly.

Although the results presented in this paper are promising in context of typical transverse profile measurement, there is still work to do towards theoretical modeling (in line with~\cite{Bagiyan}) at least for simpler surface structures~\cite{Rutkowski}, and experimental verification of the results as a function of surface roughness and irradiation angle on light yield. An evaluation of the precision of the profiles measured with unpolarized component of the transition radiation in context of scattering from rough surfaces and changes in surface due to effect of heavy ion bombardment is also required. This is relevant for detailed  beam profile measurements for high intensity beams such as halo measurements or measurements of very narrow beams ($< 1$ mm). There are other potential applications e.g. if rough surfaces and multiple radiators could produce more than one photon per ion for high energy beams; non destructive OTR based particle counters for high energy beams can be foreseen. 

\section{Acknowledgment}
 C. Andre, A. Ahmed, P. Forck, M. Mueller and S. Udrea are acknowledged for multiple discussions and help in mechanical set-up. GSI Target laboratory colleagues for providing the stripper foils and GSI Material science department colleagues for providing Glassy carbon target for these experiments are also gratefully acknowledged. Finally, we thank the UNILAC operations crew for their support during the experiments.

\section{Appendix}

\subsection{Transmission and conversion with the optical system onto the CCD}\label{App:OpticalSystem}

\begin{figure}[h]
\centering  
    \includegraphics[width=0.75\textwidth]{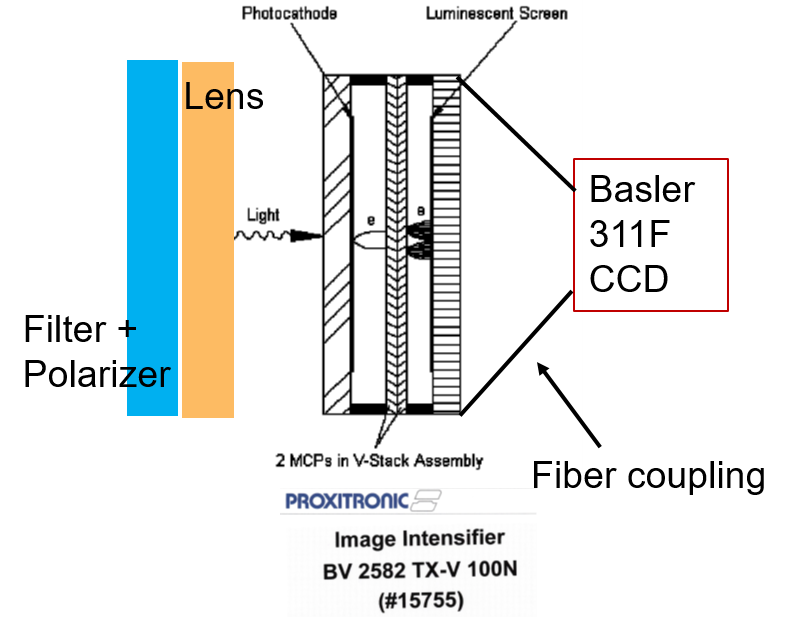}
    \caption{Optical system}
    \label{fig:opt}
\end{figure}

Figure~\ref{fig:opt} shows the components of the optical system. The optical system consists of few parameters, a) Quantum efficiency of the photocathode, b) MCP voltage and resulting electron multiplication (gain) and c) Phospor converting the electrons back to photons which are detected on the CCD. 
Photocathode quantum efficiency spectral response for the given system is presented in the table below: 
\begin{center}
\begin{tabular}{ |c|c|c| } 
 \hline
 wavelength (nm) & sensitivity (mA/W) & QE (\%) \\ 
200 & 22.0 & 13.7 \\ 
220 & 23.8 & 13.4 \\ 
240 & 25.8 & 13.3 \\ 
260 & 31.8 & 15.1 \\
280 & 37.5 & 16.6 \\ 
300 & 31.5 & 13.0 \\ 
320 & 34.5 & 13.3 \\ 
340 & 39.0 & 14.2 \\
360 & 41.6 & 14.3 \\ 
380 & 45.5 & 14.8 \\ 
400 & 49.0 & 15.2 \\ 
450 & 51.0& 14.1 \\
514 & 39.0 & 9.4 \\
650 & 21.0 & 4.0 \\ 
800 & 0.2 & 0.0 \\ 
 \hline
\end{tabular}
\end{center}
{\bf Total number of photons detected at the CCD as a function of photons entering the ICCD:}
Photocathode: QE at 400 nm = 0.15 , i.e. 0.15 el/ph. 10-15\% of the incoming photons are detected. If the MCP gain of 1.2 kV is applied; 3000 el/el result; similary for 1.6 kV = 300000 el/el. (See manual/slides). Phosphor 46 provides 90 ph/el at 6 KV (default voltage between MCP and Phosphor). Thus the total input to output photon conversion at MCP gain 1.2 kV($\approx 4\cdot 10^4$) and 1.6 kV $(\approx 4\cdot 10^6)$ ph/ph.

The quantum efficiency of the Basler 311F CCD camera is not yet clear. The default pixel values without beam was $500\pm60$ (in 2 byte mode). Thus there are 7 effective bits allowing a dynamic range of ~40 dB (factor 100).
The intensifier was aimed to be used in photon counting mode, i.e. a high gain around 1.4 kV was set where each photon detected on the photocathode resulted in $10^6$ photons on the CCD making a "blob". This set-up is only appropriate for low photon scenario ($\leq 50$ photons per image) at smaller values of $\psi$. We noticed CCD saturation for individual images for Aluminum target at $\psi = 10\degree$ and $20 \degree$ during our measurements with 6.0 MeV/u $5\cdot 10^9$ $Ca^{10+}$ ions.  

\subsection{CCD count fluctuations}\label{App:ShotNoise}
An independent way of estimating the number of photons reaching the CCD is to evaluate the fluctuations in the number of counts for consecutive images under the same beam and target conditions under constant beam conditions. These type of discrete event number related fluctuations are referred to as "shot" noise in electronics literature and is modelled by a Poisson process. 

For a Poisson process with an average of $N$ events in a given time interval, the shot noise is expected to be $\sqrt{N}$.
Beam current is measured independently using a current transformer and was $40\pm1$ $\mu$A for the discussed data. The beam current fluctuations contribute very weakly to the count fluctuations since $\frac{\sigma_I}{<I>} \ll \frac{\sigma_{counts}}{<counts>}$. Based on this data itself, it is not clear if the fluctuations are dominated by the noise in image intensification process or the photons generation process, however it does allow an estimate on the lower bound on the number of events reaching the CCD.
Assuming that the fluctuation is count rates is dominated by the shot noise of the generated photons, the average number of events or generated photons can be estimated as
\begin{align}
N_{ph,average}=\big(\frac{<counts>}{\sigma_{counts}}\big)^2
\end{align}
\begin{figure}[h]
        \subfigure[Permittivity]{\label{fig:otspeca}
            \includegraphics[width=0.475\textwidth]{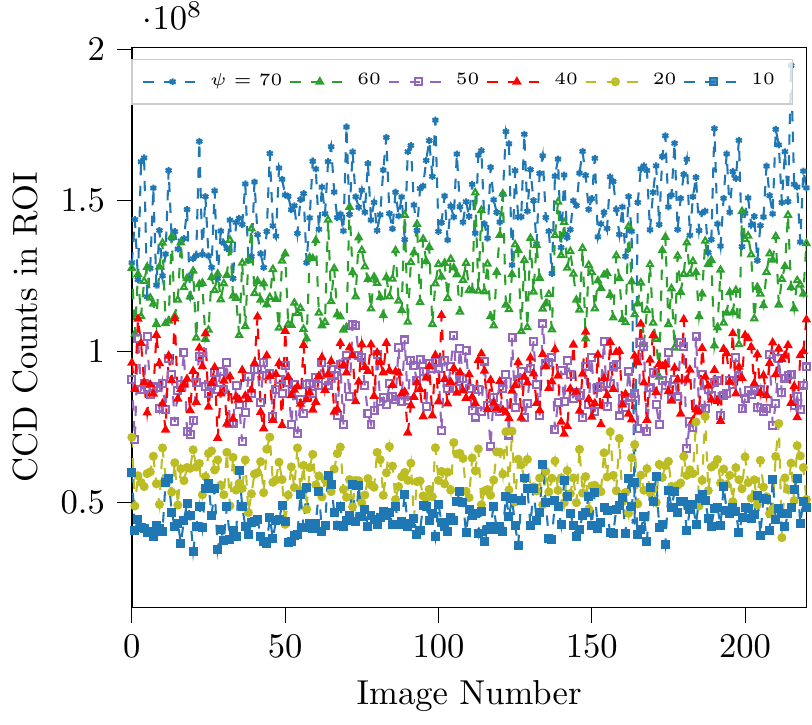}}
        \subfigure[$\psi = 10$]{\label{fig:otspecb}
            \includegraphics[width=0.475\textwidth]{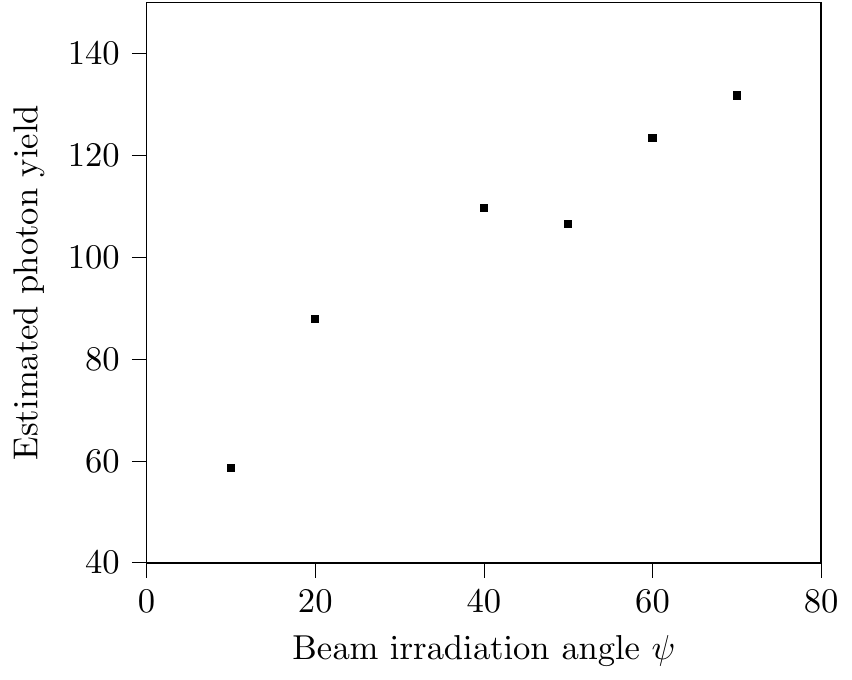}}
    \caption{CCD counts from Aluminum target for various irradiation angles when the polarizer angle is set to 0$\degree$.}
    \label{fig:shot-noise}
\end{figure}
Figure~\ref{fig:shot-noise} (a) shows the fluctuations in the CCD count for irradiation of Aluminum target for several incidence angles. The estimated photon counts are shown in Figure~\ref{fig:shot-noise} (b).
We have to note that, the role of background subtraction is crucial for this process since this is an absolute photon number estimate. For $\cdot10^9 Ca^{10+}$ beam with $\beta = 0.11.$, we estimate generation of 1000 photons for Aluminium target at $\psi=30 \degree$ (Fig.~\ref{fig:photon_count}. As discussed in~\ref{App:OpticalSystem}, the photocathode has an average quantium efficiency of 15\%, and the optical system should register a maximum of $\approx 150$ photons per image from a smooth target. On the other hand, the photon estimate from the measurements is from a rough Aluminum target, and the maximum intensity (140 photons) is measured at $\psi=70$.

In comparison to photon yields above, the beam induced fluorescence monitor provides roughly $\approx 50$ times lower number of photons when scaled for the same beam conditions~\cite{C_andre_bif}. As discussed above, 500 events or photons are sufficient to provide a reliable 1 D profile for an image intensified system.
\subsection{OTR spectra for Iron, Aluminum and Glassy carbon}\label{App:spectra}
It is clear from Eq.~\ref{eq:tr} that the transition radiation spectral intensity is a rather complex function of material permittivity, angle of irradiation and angle of observation. For low betas, e.g.  $(\beta <<1)$, it reduces to a simpler form.
\begin{equation}\label{eq:tr2}
\frac{dI_\parallel(n,\omega,\theta,\phi,\psi)}{d\Omega d\omega}= \frac{Z^2e^2\beta^{2}\cos^2{\psi}}{4\pi^3\epsilon_0 c}\Bigg|\frac{\sin{\theta}\cos{\theta}(\epsilon-1)}{\epsilon \cos{\theta}+\sqrt{\epsilon-\sin^2{\theta}}}\Bigg|^2
\end{equation}

\begin{figure}
        \centering
        \subfigure[Permittivity]{\label{fig:otspeca}
            \includegraphics[width=0.475\textwidth]{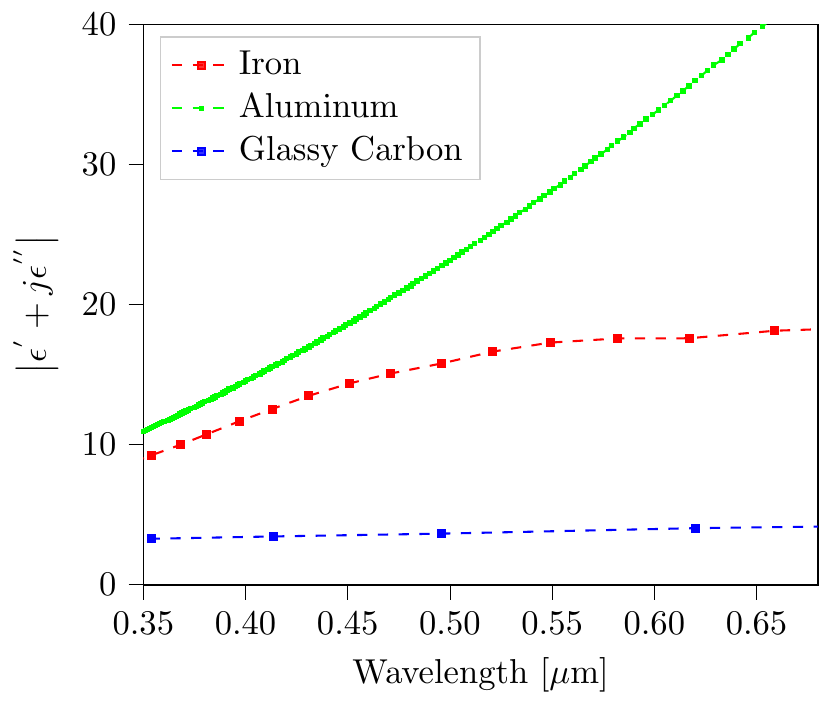}}
        \subfigure[$\psi = 10$]{\label{fig:otspecb}
            \includegraphics[width=0.475\textwidth]{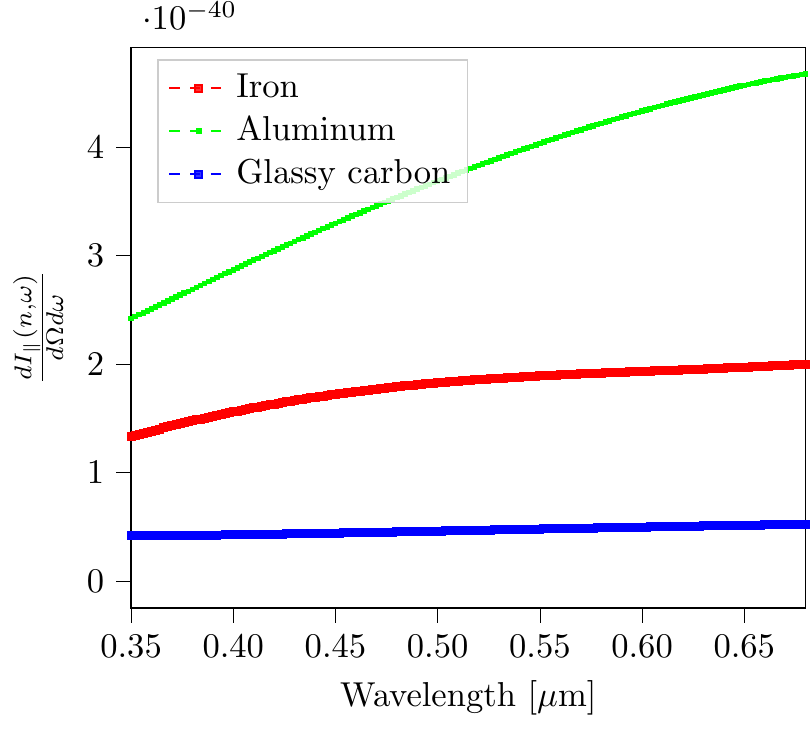}}
        \vskip\baselineskip
        \subfigure[$\psi = 40$]{\label{fig:otspecc}
            \includegraphics[width=0.475\textwidth]{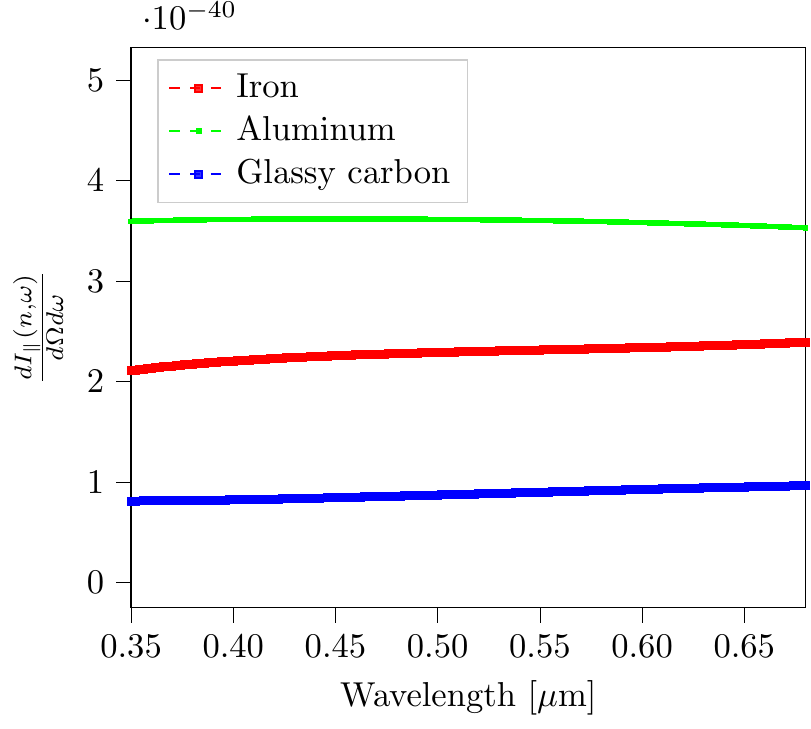}}
        \subfigure[$\psi = 70$]{\label{fig:otspecd}
            \includegraphics[width=0.475\textwidth]{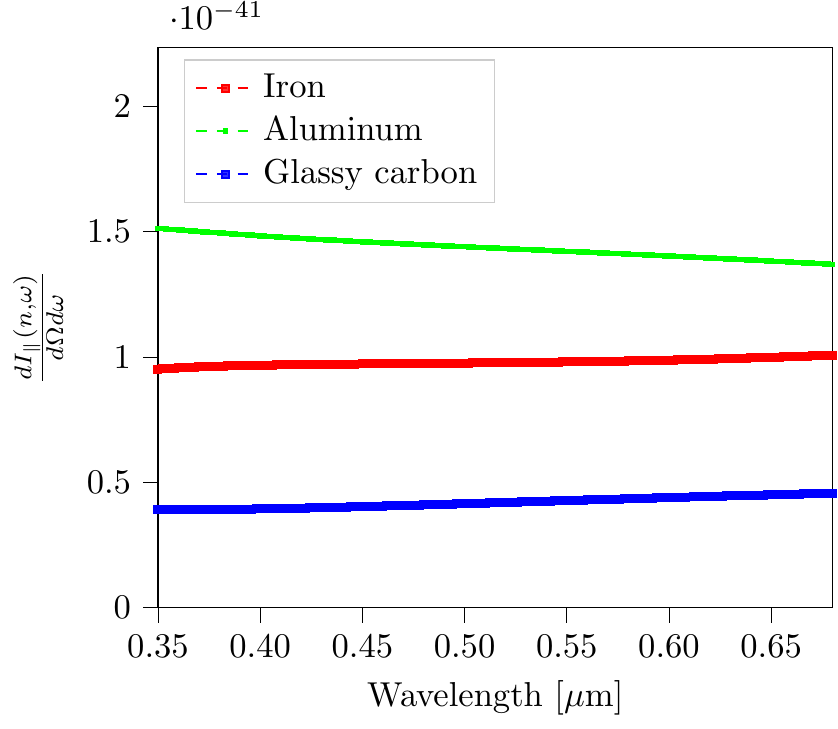}}
        \caption[]
        {Optical transition radiation spectra. (a) Shows the variation in absolute value of epsilon in visible range. (b-d) Calculated radiated spectral intensity for different irradiation angles and the observation angle is $\theta = 90-\psi$.} 
        \label{fig:TRSpectra}
    \end{figure}

Let us consider two extremes of observation angle $\theta$ where radiation intensity becomes lower due to $\cos{\theta}\sin{\theta}$ dependence, however the effect of material property dependence is easier to highlight. At any given irradiation angle $\psi$, as the angle of observation $\theta \to \pi/2$, the spectral intensity $\frac{dI_\parallel(n,\omega)}{d\Omega d\omega} \approx |\epsilon-1|$. Towards the other extreme as $\theta \to 0$, $\frac{dI_\parallel(n,\omega)}{d\Omega d\omega} \approx |\frac{\sqrt{\epsilon}-1}{\sqrt{\epsilon}}|^2$, which for sufficiently large values of $\epsilon$ approaches 1, and will result in a flat spectra. Thus depending on the angle of observation, the optical transition spectra will have signatures of the frequency dependence of absolute value of permittivity.  This is highlighted in Fig.~\ref{fig:TRSpectra}, where spectral intensity of p-polarized component is shown for a particle pf unit charge travelling with $\beta = 0.11$ and incident on the three target materials Aluminum, Iron and Glassy carbon. The frequency dependence of permittivity values are acquired from ~\cite{Cheng,Christy,Williams} respectively.
\subsection{Expected photon number for PEC at normal incidence}\label{App:NphNormPEC}
Setting $\epsilon = 10^{12}$ (to emulate a perfect electric conductor) and $\psi = 0$ we can compare photon number \eqref{eq:tr_par1} derived from spectral intensity \eqref{eq:tr} for oblique case with the result obtained using spectral energy (\cite{Ginzburg1})
\begin{align}\label{eq:specEnergyPEC}
    \frac{dE}{d\omega} = \frac{Z^2 e^2}{4\pi^2 \epsilon_0 c }\left( \frac{1+\beta^2}{2\beta}\ln\left(\frac{1+\beta}{1-\beta}\right) -1 \right)
\end{align}
which is valid for normal incidence and PEC. In the latter case the total photon number is given as
\begin{align}\label{eq:NphPECnormal}
    N_\text{ph,PEC,$\perp$} = \frac{N_{ion}}{\hbar}\ln\left(\frac{\omega_2}{\omega_1}\right)  \frac{dE}{d\omega} 
\end{align}
For $N_{ion} = 5\cdot 10^9$ ions with charge state $Z=10$ travelling with $\beta = 0.11$ and in the wavelength range $\lambda \in [350, 650]$~nm one obtains  $1.166\cdot 10^7$ photons using \eqref{eq:NphPECnormal}. This gives a scaling of $N_{ph,t} = 2\cdot10^{-3}\beta^2Z^2N_{ions}$ inline with Eq.~\ref{eq:tot_photons}.
\subsection{Number of photons needed for profile measurement}\label{App:HowManyPhotons}

Modern detectors are capable of detecting every photon which are registered as an electron on the input of an MCP or PMT due to their large amplification. Therefore the question of the number of photons required to form a profile image has moved towards the source~\cite{Johnson}; how many photons should be emitted to obtain a reasonable beam profile image?
The answer to this question depends on the resolution requirements. If we want a 2-D beam image with a cross section of 15 mm $\times$ 15 mm with 0.5 mm spatial resolution and an amplitude resolution of 3\% or 30 levels (5 bits). The total degrees of freedom for this problem are $N_{df}=30\times30\times30 = 27000$. Since the absence or presence of a photon convey information to a binary bit, the total number of photons required is $N_{ph,t,2D}= N_{df}/2 = 13500$.

For obtaining 2 1-D projections of the profile, the requirements are much meagre, $N_{ph,t,1D}= N_{ph,t,1D}/15 \approx 1000$. Such a requirement can easily be fulfilled for the typical intensities at heavy ion and proton linacs.
\subsection{Long term irradiation}
\begin{figure}[h]
\centering  
    \includegraphics[width=0.99\textwidth]{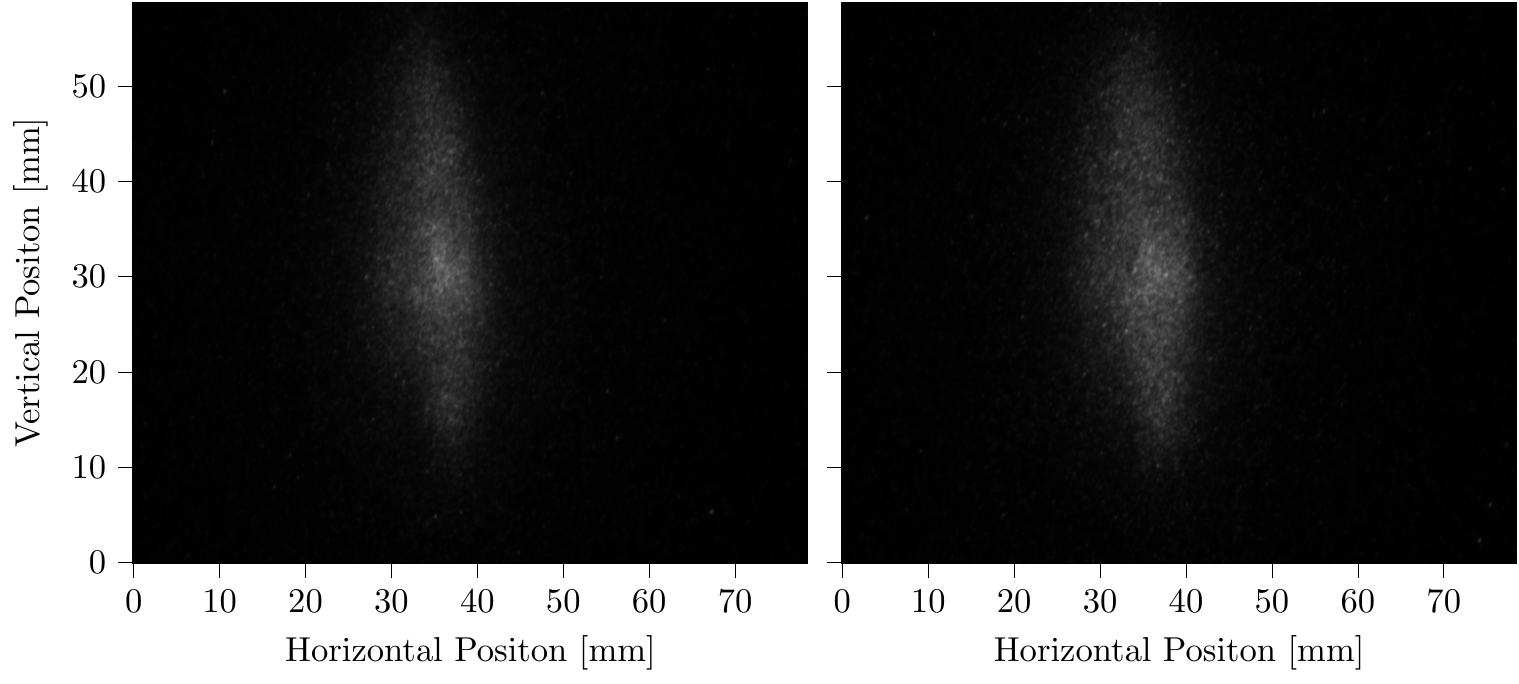}
    \caption{Left: Image at the start of half an hour irradiation. Right: Image after half an hour of irradiation.}
    \label{fig:IrrStudy}
\end{figure}

Figure~\ref{fig:IrrStudy} shows two images spaced by an interval of half an hour of high intensity  beam irradiation on Gold target. Approximately $\approx 10^{13}$ $Bi^{26+}$ ions for irradiated on the target in this time interval. No change in the measured beam profile is observed.
\end{document}